# On the Complexity of Finding Narrow Proofs

Christoph Berkholz

June 18, 2018

We study the complexity of the following "resolution width problem": Does a given 3-CNF have a resolution refutation of width $k$? We prove that the problem cannot be decided in time $O(n^{(k-3)/12})$. This lower bound is unconditional and does not rely on any unproven complexity theoretic assumptions. The lower bound is matched by a trivial upper bound of $n^{O(k)}$.

We also prove that the resolution width problem is EXPTIME-complete (if $k$ is part of the input). This confirms a conjecture by Vardi, who has first raised the question for the complexity of the resolution width problem. Furthermore, we prove that the variant of the resolution width problem for regular resolution is PSPACE-complete, confirming a conjecture by Urquhart.

## 1. Introduction

Resolution is a well-known and intensively studied proof system to detect the unsatisfiability of a given formula in conjunctive normal form (CNF). Starting with the clauses from the CNF formula one iteratively derives new clauses using only one simple rule: The resolution rule takes two clauses $\gamma \cup \{X\}$, $\delta \cup \{\neg X\}$ and resolves $\gamma \cup \delta$. The given CNF formula is unsatisfiable if, and only if, the empty clause can be derived. Despite its simplicity resolution has been found many applications in practical SAT solving. Most state-of-the-art SAT solvers try to find resolution refutations.

One natural complexity measure for resolution is the *length* of a refutation. This measure is also important for resolution based satisfiability testing since the running time of that SAT solvers is lower bounded by the length of the underlying resolution refutation. Haken [11] proved the first superpolynomial lower bound on the length of resolution refutations for the pigeon hole principle. Several improvements and length lower bounds for other combinatorial principles followed. A second complexity measure is the *width* of a resolution refutation, which is the size of the largest clause in the refutation. Ben-Sasson and Widgerson [6] underlined its importance by showing that every length $S$ resolution refutation of an $n$-variable 3-CNF formula can be transformed to a refutation of width at most $O(\sqrt{n \log S})$. Hence, if a 3-CNF formula has a "short" (subexponential) refutation, then it has also a "narrow" refutation of sublinear width. This fact enabled them to rederive essentially all previous known exponential length lower bounds by proving linear width lower bounds. Furthermore, they proposed a simple dynamic algorithm that searches for a refutation of smallest width. This heuristics was already known before and dates back to Galil [9]. It proceeds in a very simple way:

$i \leftarrow 0$.
**repeat**
  $i \leftarrow i + 1$.
  Derive all clauses of width at most $i$.
**until** the empty clause has been derived.

Since on $n$ variables there are at most $O(n^k)$ clauses of width $k$, the algorithm terminates after $n^{O(w)}$ steps, where $w$ is the smallest width of a resolution refutation of $\Gamma$. To estimate the running time of this procedure on a given instance, one needs to solve the following decision problem.

| Resolution width problem | |
|---|---|
| *Input*: | A 3-CNF formula $\Gamma$ and an integer $k$. |
| *Question*: | Does $\Gamma$ have a resolution refutation of width at most $k$? |

The algorithm above solves this problem within exponential time by deriving all clauses of width at most $k$. Our first theorem states that this problem cannot be solved within polynomial time.

**Theorem 1.** *The resolution width problem is complete for* EXPTIME.

Motivated by an EXPTIME-completeness result for the k-consistency heuristics for general CSP [16], Vardi raised the question for the complexity of the resolution width problem and conjectured that it is EXPTIME-complete. In 2006, Hertel and Urquhart [12] claimed to have solved the problem, but later retracted their claim [13]. Nordström mentions it as an open problem in his recent survey [17]. A related problem is the *regular resolution width problem* that asks whether or not there exists a *regular* resolution refutation of width at most $k$. Urquhart stated its complexity as open problem and conjectured it to be PSPACE-complete [20]. We settle this conjecture as well.

**Theorem 2.** *The regular resolution width problem is complete for* PSPACE.

For more motivation of the above theorems we refer to Chapter 7 of Hertel's dissertation [14] that also discusses quite a few interesting consequences. If an unsatisfiable 3-CNF formula can be refuted by a constant width resolution refutation, then the algorithm above recognizes its unsatisfiability within polynomial time. Thus, searching for width-$k$ refutations may serve as polynomial time heuristics for determining unsatisfiability. On the other hand the degree of the polynomial depends on $k$ and it is natural to ask whether this is necessary. That is, can the following decision problem be solved in, say, quadratic time?

| Resolution width-$k$ problem | |
|---|---|
| *Input*: | A 3-CNF formula $\Gamma$. |
| *Question*: | Does $\Gamma$ have a resolution refutation of width at most $k$? |

The existence of an $O(2^k\|\Gamma\|^2)$ time, hence quadratic, algorithm for the resolution width-$k$ problem would be consistent with Theorem 1 and with our previous knowledge. Our third theorem rules out this possibility in a very strong manner.

**Theorem 3.** *For every integer $k \geq 15$, the resolution width-k problem can not be decided in time $O(\|\Gamma\|^{\frac{k-3}{12}})$ for a given 3-CNF formula $\Gamma$ on multi-tape Turing machines.*

Note that this lower bound is unconditional because it is ultimately obtained from the deterministic time hierarchy theorem. The simple algorithm above computes a resolution refutation of width at most $k$, provided there is one, in time $\|\Gamma\|^{O(k)}$. Hence, this theorem also states that there is no significant better way to decide the existence of a width-k refutation than exhaustively searching for it. The proof of Theorem 3 also settles the parameterized complexity of the resolution width problem:

**Corollary 4.** *Parameterized by the width $k$, the resolution width problem is complete for XP.*

This corollary adds one more natural problem to the short list of XP-complete problems. However, Theorem 3 is stronger in the sense that XP-completeness does not rule out the possibility of time $O(n^{\log\log k})$ algorithms.

As mentioned above, every width-$k$ refutation has length at most $O(n^k)$ where $n$ is the number of variables in the 3-CNF formula and it is an intriguing question if this bound is sharp. We prove for every constant $k$ a near optimal lower bound by explicitly constructing a family $\{\Gamma_n^k\}_{n=1}^{\infty}$ of 3-CNF formulas with $O(n)$ variables that can be refuted in width-$k$ resolution, but for which every width-$k$ resolution refutation has length at least $\Omega(n^{k-1})$. On the other hand, $\Gamma_n^k$ can be refuted by a treelike resolution refutation of width $k+1$ and constant length (depending on $k$). Thus, the refutation of smallest width is by means longer than the shortest one. Such a trade-off was unknown before and relates to the work of [3] and open problems in [18] (see also [17, (Chapter 6)] for further discussion).

**Theorem 5.** *For every fixed integer $k \geq 3$ there is a family of unsatisfiable 3-CNF formulas $\{\Gamma_n^k\}_{n=1}^{\infty}$ with $O(n)$ variables, $O(n^2)$ clauses and minimal refutation width $k$ such that the following holds:*

- *Every width-$k$ resolution refutation of $\Gamma_n^k$ has length at least $\Omega(n^{k-1})$.*
- *There is a width-$(k+1)$ treelike resolution refutation of $\Gamma_n^k$ of length $O(1)$.*

The three computational lower bounds stated above are obtained by essentially one reduction from the combinatorial KAI-game [15] to a restricted variant of the existential pebble game that characterizes resolution width [2]. Our proofs built on earlier work by Kolaitis and Panttaja [16] and recent work by the author of this paper [7] on the complexity of existential pebble games. We introduce both games and state the reduction in the next section. Section 3 summarizes the proof techniques and outlines the reduction, the details of the reduction are given in Section 4 and 5. In Section 6 we sketch the lower bound on the length of width-k resolution refutations and present a full proof in the appendix.

# 2. Definitions and Proof of the Main Theorems

## 2.1. A Game Characterization of Resolution Width

A *literal* is either a Boolean variable $X$ or its negation $\neg X$. A *clause* $\gamma$ is a disjunction of literals and the *width* of a clause is the number of literals in it. A CNF formula $\Gamma$ is a conjunction of clauses and a $d$-CNF formula is a CNF formula that contains only clauses of width at most $d$. It is common to view clauses as sets of literals and formulas as sets of clauses. Resolution is a well-known calculus for proving the unsatisfiability of a given CNF formula. The *resolution rule* on $X$ takes two clauses $\gamma \cup \{X\}$ and $\delta \cup \{\neg X\}$ and derives the *resolvent* $\gamma \cup \delta$. A *resolution derivation* of a clause $\gamma$ from a CNF formula $\Gamma$ is a sequence of clauses $(\gamma_1, \ldots, \gamma_n)$ such that $\gamma = \gamma_n$ and every clause $\gamma_i$ is either contained in $\Gamma$ or a resolvent of two preceding clauses. A *resolution refutation* is a resolution derivation of the empty clause.

The *length* of a resolution derivation is the number of clauses it contains and the *width* of a resolution derivation is the maximum width over all clauses in that derivation. A resolution derivation of $\gamma$ can also be viewed as a directed acyclic graph (dag) where the nodes are labeled with the clauses from the derivation, one node of in-degree 0 is labeled with $\gamma$ and all nodes of out-degree 0 are labeled with clauses from $\Gamma$. There is one arc from $\delta$ to $\gamma_1$ and one arc from $\delta$ to $\gamma_2$ if $\delta$ is the resolvent of $\gamma_1$ and $\gamma_2$. The depth of a resolution derivation of $\gamma$ from $\Gamma$ is number of arcs on the longest directed path in the corresponding dag. A resolution derivation is *regular* if on every path from the root to the leafs in the associated dag no variable has been used twice by the resolution rule.

A *partial assignment* is a partial mapping $p$ from the Boolean variables to $\{0, 1\}$. The Boolean existential $(k+1)$-pebble game introduced by [2] works with these partial assignments and is designed to simulate width-$k$ resolution. This game can be seen as a special case of the model-theoretic existential $(k+1)$-pebble game. On the other hand it is quite similar to Pudlak's Prover-Delayer game for resolution [19] if one bounds the size of the so-called record. For abbreviation we call the Boolean existential $(k+1)$-pebble game "width-$k$ game" here. The game is played by two players, called Spoiler and Duplicator, and the positions of the game are partial assignments of domain size at most $k+1$. The game starts with the empty assignment. In each round, Spoiler asks Duplicator for the assignment of a variable $X$ and Duplicator has to answer with either $X \mapsto 0$ or $X \mapsto 1$. Spoiler can store at most $k+1$ variables and its assignments, but he can delete information at any time. After Spoiler has stored the $(k+1)$st assignment, he is forced to delete at least one assignment before doing anything else. Spoiler wins the game if he can reach an assignment that falsifies a clause from $\Gamma$ and Duplicator wins the game if she has a strategy such that Spoiler can never reach such a position. For illustration we also view a partial assignment $p$ of domain size $l$ as a set of $l$ pebbles marked with 0 or 1 and lying on the variables $\mathrm{Dom}(p)$. In [2] it was shown that Spoiler wins the width-$k$ game on $\Gamma$ if, and only if, $\Gamma$ has a resolution refutation of width at most $k$. The next lemma relates also the depth of a width-$k$ refutation to the number of rounds in the width-$k$ game.

**Lemma 6.** *Spoiler wins the width-$k$ game on $\Gamma$ within $d$ rounds if, and only if, $\Gamma$ has a resolution refutation of width at most $k$ and depth at most $d$.*

*Proof.* We first show how a width-$k$ resolution refutation leads to a winning strategy for Spoiler. We can identify every clause $\gamma$ of width $l$ with the unique partial assignment of domain size $l$ that falsifies it. For example, the clause $\{X, \neg Y, Z\}$ is falsified by the partial assignment $\{X \mapsto 0, Y \mapsto 1, Z \mapsto 0\}$. Spoiler plays along the arcs in the resolution dag from the empty clause to some clause in $\Gamma$ and always stores the assignment that falsifies the current clause (hence this assignment has domain size at most $k$). First, the game starts with the empty assignment that corresponds to the empty clause in the derivation. If the current clause is derived from $\gamma_1 \cup \{X\}$ and $\gamma_2 \cup \{\neg X\}$ via resolving on $X$, then Spoiler asks for $X$. Depending on Duplicators choice, he walks to either of the two parents and deletes assignments that are not related to the new clause. Finally, he reaches an assignment that falsifies a clause from $\Gamma$ and thus he wins. Since he follows a path from the root to the leafs in the dag, the number of rounds is bounded by the depth of the refutation.

In an analog way one can develop a resolution refutation of width at most $k$ from a winning strategy for Spoiler in the width-$k$ game. In order to do this we first construct a resolution refutation that also uses the *weakening rule* that derives a clause $\gamma$ from a clause $\delta \subset \gamma$. A resolution refutation with weakening can easily be transformed to a standard resolution refutation without increasing length, width and depth. The refutation we construct uses the clauses that are falsified by the current assignment, if the domain size is less than $k+1$. For every partial assignment of domain size $k+1$ occurring in the strategy, we consider the clause that relates to the corresponding partial assignment *after* Spoiler was forced to delete one variable. Deleting assignments in Spoilers strategy corresponds to weakening. If Spoiler asks for $X$ this essentially corresponds to resolving on $X$, but we have to be a little bit more precise here. Let $\gamma$ be the clause that relates to the current assignment (that falsifies it) and $X$ be the variable Spoiler asks for. If $|\gamma| < k$, then $\gamma$ is obtained from $\gamma \cup \{X\}$ and $\gamma \cup \{\neg X\}$ via resolving on $X$. If $|\gamma| = k$, let $\gamma_1 \subset \gamma \cup \{X\}$ and $\gamma_2 \subset \gamma \cup \{\neg X\}$ be the clauses obtained after Spoiler was forced to delete at least one assignment. Now it holds that (1) $\gamma$ is (a weakening of) $\gamma_1$ or (2) $\gamma$ is (a weakening of) $\gamma_2$ or (3) $X \in \gamma_1$ and $\neg X \in \gamma_2$ and $\gamma$ is (a weakening of) the resolvent of $\gamma_1$ and $\gamma_2$. Since every play of the game relates to a path from the empty clause to some clause in $\Gamma$ in the resolution-dag we get a width-$k$ resolution refutation of depth at most $d$ (after getting rid of the weakening). $\square$

A slight modification of the width-$k$ game yields an appropriate game to characterize regular resolution refutations of width at most $k$ [14]. The *regular width-$k$ game* proceeds as the width-$k$ game with the restriction that Spoiler is not allowed to ask for a variable twice. The following lemma is a straightforward adaptation of Lemma 6.

**Lemma 7.** *Spoiler wins the regular width-$k$ game on $\Gamma$ within $d$ rounds if, and only if, $\Gamma$ has a regular resolution refutation of width at most $k$ and depth at most $d$.*

## 2.2. The Pebble Games of Kasai, Adachi and Iwata

An instance of the KAI-game [15] is a tuple $(U, R, \mathfrak{s}, \theta)$ where $U$ is the universe, $R = R' \times \binom{[k]}{2}$ with $R' \subseteq U^3$ the set of rules, $\mathfrak{s}\colon [k] \to U$ the start position and $\theta \in U$ the goal. We let $[k]$ be the set of $k$ pebbles in the game. A rule is of the form $(u, v, w, c, d)$, with $c \neq d$, $u \neq v \neq w \neq u$ and the intended meaning that if pebble $c$ is on $u$ and pebble $d$ is on $v$ and there is no pebble on $w$, then one player can move pebble $c$ from $u$ to $w$. This is a slight more wasteful notion as originally used in [15], where the set of rules is $R' \subseteq U^3$, but it is useful in our reduction to specify the pebbles $c$ and $d$ in the rules. A *position* of the KAI-game is an injective mapping $\mathfrak{p}\colon [k] \to U$. A rule $r = (u, v, w, c, d) \in R$ is *applicable* to a position $\mathfrak{p}$ if $\mathfrak{p}(c) = u$, $\mathfrak{p}(d) = v$ and $\mathfrak{p}(z) \neq w$ for all $z \in [k]$. Furthermore, $r(\mathfrak{p})$ denotes the position defined as $r(\mathfrak{p})(c) = w$ and $r(\mathfrak{p})(z) = \mathfrak{p}(z)$, for all $z \in [k] \setminus \{c\}$. If $r$ is applicable to $\mathfrak{p}$, then $r(\mathfrak{p})$ is the position that occurs after applying $r$ to $\mathfrak{p}$. The set of all rules in $R$ applicable to a position $\mathfrak{p}$ is denoted by $\mathrm{appl}(\mathfrak{p})$ and $T_r(\mathfrak{p}) \subseteq [k]$ denotes the set of pebbles $i$ such that $\mathfrak{p}(i)$ contradicts the applicability condition of rule $r$: $T_{(u,v,w,c,d)}(\mathfrak{p}) := \{i \in [k] \mid (i = c \text{ and } \mathfrak{p}(i) \neq u) \text{ or } (i = d \text{ and } \mathfrak{p}(i) \neq v) \text{ or } \mathfrak{p}(i) = w\}$.

The KAI-game is played by two players and proceeds in rounds. In the first round Player 1 starts with position $\mathfrak{s}$ and chooses a rule $r \in \mathrm{appl}(\mathfrak{s})$, the new position is $\mathfrak{p} = r(\mathfrak{s})$. In the next round Player 2 chooses a rule $r \in \mathrm{appl}(\mathfrak{p})$ and applies it to $\mathfrak{p}$. Then it is Player 1's turn and so on. Player 1 wins the game if he reaches a position $\mathfrak{p}$, where $\mathfrak{p}(z) = \theta$ for one $z \in [k]$ (that is called a *winning position*) or where Player 2 is unable to move. Player 2 wins if she has a strategy ensuring that Player 1 cannot reach such a position. The next definition formalizes winning strategies for Player 2, they contain a set of positions $\mathcal{K}_1$ where it is Player 1's turn and a set of positions $\mathcal{K}_2$ where it is Player 2's turn and a function $\kappa$ that tells Player 2 which rule to choose next.

**Definition 8.** A *winning strategy* for Player 2 in the KAI-game on $G = (U, \{r_1, \ldots, r_m\}, \mathfrak{s}, \theta)$ is a triple $\mathcal{K} = (\mathcal{K}_1, \mathcal{K}_2, \kappa)$ where $\mathcal{K}_1 \subseteq \{\mathfrak{p} \mid \mathfrak{p}\colon [k] \to U\}$ and $\mathcal{K}_2 \subseteq \{\mathfrak{p} \mid \mathfrak{p}\colon [k] \to U \setminus \{\theta\}\}$ are sets of positions and $\kappa\colon \mathcal{K}_2 \to [m]$ is a mapping such that the following holds:

- $\mathfrak{s} \in \mathcal{K}_1$.
- For every $\mathfrak{p} \in \mathcal{K}_1$ and every $r_i \in \mathrm{appl}(\mathfrak{p})$: $r_i(\mathfrak{p}) \in \mathcal{K}_2$.
- For every $\mathfrak{p} \in \mathcal{K}_2$: $r_{\kappa(\mathfrak{p})} \in \mathrm{appl}(\mathfrak{p})$ and $r_{\kappa(\mathfrak{p})}(\mathfrak{p}) \in \mathcal{K}_1$.

In the *$k$-pebble KAI-game* the instances are required have to exactly $k$ pebbles (as indicated by the start position). The underlying directed graph of a KAI-game instance $G = (U, R, \mathfrak{s}, \theta)$ consists of the node set $U$ and arcs $(u, w)$ and $(v, w)$ for every rule $(u, v, w, c, d) \in R$. An instance of the KAI-game is *acyclic* if its underlying directed graph is acyclic and the *acyclic KAI-game* is the KAI-game restricted to acyclic instances. The next theorem from [15] addresses the complexity of deciding which player wins the (acyclic) KAI-game.

**Theorem 9.** *Determining the winner in the KAI-game is complete for* EXPTIME *and determining the winner in the acyclic KAI-game is complete for* PSPACE*.*

It can be decided in time $n^{O(k)}$ if Player 1 has a winning strategy in the $k$-pebble KAI-game on $G$, thus this problem is in PTIME for every fixed $k$. Theorem 10 below states a corresponding lower bound. It was proven in [1] by simulating a deterministic multi-tape Turing machine of running time $n^k$ within the $k'$-pebble KAI-game so that the machine accepts if, and only if, Player 1 wins the KAI-game. The lower bound then follows from the time hierarchy theorem, that states that Turing machines of running time $n^k$ cannot be simulated in time $n^{k-\varepsilon}$.

**Theorem 10.** *For every $\varepsilon > 0$, determining the winner in the $k$-pebble KAI-game is not in* $\mathrm{DTIME}(n^{\frac{k-1}{4}-\varepsilon})$.

## 2.3. Proof of the Main Theorems

We write (*regular*) *width game* to denote that the parameter $k$ is given as part of the input. We now prove the computational lower bounds, using the reductions stated in the next two lemmas. The main lemmas itself are proven at the end of Section 5.

**Lemma 11** (First Main Lemma)**.** *There is a* LOGSPACE*-reduction from the KAI-game to the width game and from the acyclic KAI-game to the regular width game.*

*Proof of Theorem 1.* It is easy to see that the resolution width problem is in EXPTIME by iteratively resolving all clauses of width at most $k$. Since determining the winner in the KAI-game is EXPTIME-hard (Theorem 9) it is EXPTIME-hard to determine the winner in the width game by Lemma 11. Hence, the resolution width problem is complete for EXPTIME. □

*Proof of Theorem 2.* Spoiler has a forced win in the regular width game if, and only if, he can win the game within $|\mathrm{Var}(\Gamma)|$ steps. Thus, an alternating Turing machine can decide if Spoiler can win the game in polynomial time. By APTIME=PSPACE [8] we get that the regular resolution width problem is in PSPACE. Since the acyclic KAI-game is PSPACE-hard (Theorem 9) and there is a LOGSPACE-reduction from the acyclic KAI-game to the regular width game (Lemma 11) it follows that the regular resolution width problem is complete for PSPACE. □

**Lemma 12** (Second Main Lemma)**.** *There is a reduction from the $k$-pebble KAI-game to the width-$(k+1)$ game that computes for every instance $G$ of size $\|G\|$ a 3-CNF formula $\Gamma(G)$ such that the following holds.*

- *Player 1 has a winning strategy in the $k$-pebble KAI-game on $G$ if, and only if, Spoiler has a winning strategy in the width-$(k+1)$ game on $\Gamma(G)$.*
- *$\Gamma(G)$ contains $O(\|G\|^3)$ clauses and $O(\|G\|^2)$ variables.*
- *The reduction is computable in* $\mathrm{DTIME}(\|G\|^3)$.

*Proof of Theorem 3.* Let $k \geq 15$ be a fixed integer. Assume that $\mathbb{A}$ is an algorithm that determines the winner of the width-$k$ game on $\Gamma$ in time $O(\|\Gamma\|^{\frac{k-3}{12}})$. Let $\mathbb{B}$ be the algorithm that first applies the reduction from Lemma 12 to a given instance $G$ of the $(k-1)$-pebble KAI-game and then executes $\mathbb{A}$. Since $\|\Gamma(G)\| = O(\|G\|^3)$, $\mathbb{B}$ has running time $O(\|G\|^3 + \|G\|^{\frac{k-3}{4}})$ and thus solves the $k'$-pebble KAI-game in time $O(\|G\|^{\frac{k'-2}{4}})$ for a $k' \geq 14$. This contradicts Theorem 10. □

# 3. Proof Techniques and Outline

We devise one reduction that proves both statements in Lemma 11 and a weaker form of Lemma 12 (with $\|\Gamma(G)\| = O(\|G\|^4)$) at once. With a slight modification of that reduction we obtain the bounds from Lemma 12. For the rest of the paper let $G = (U, R, \mathfrak{s}, \theta)$ with $U = [n]$, $R = \{r_1, \dots, r_m\}$, $\mathfrak{s}\colon [k] \to [n]$ and $\theta \in [n]$ be an instance of the $k$-pebble KAI-game. We construct a 3-CNF formula $\Gamma(G)$ such that the following holds.

- Player 1 has a winning strategy in the $k$-pebble KAI-game on $G$ if, and only if, Spoiler has a winning strategy in the width-$(k+1)$ game on $\Gamma(G)$.
- If $G$ is acyclic and Player 1 has a winning strategy in the $k$-pebble KAI-game on $G$, then Spoiler has a winning strategy in the regular width-$(k+1)$ game on $\Gamma(G)$.

## 3.1. Combining Strategies

In our reduction we construct the clause set $\Gamma(G)$ out of smaller clauses sets, called gadgets. The gadgets are defined on pairwise disjoint variable sets and there are additional clauses to connect these gadgets. In order to establish a winning strategy for one player, we need to combine strategies on the gadgets to a strategy on $\Gamma$. The easier part is to do that for Spoiler with a notion obtained from finite model theory [10]. We say that Spoiler *can (regularly) reach* position $p_2$ from position $p_1$ on $\Gamma$ if he has a strategy in the (regular) width-$(k+1)$ game such that starting from position $p_1$ he either wins the game or position $p_2$ occurs in the game after some finite number of rounds. We can combine such strategies to show that Spoiler can reach some position $p$ from $\emptyset$; if $p$ falsifies a clause from $\Gamma(G)$ this gives us a winning strategy for Spoiler and hence a resolution refutation. As indicated in the proof of Lemma 6 there is a tight connection between strategies for Spoiler and resolution derivations. If $|\operatorname{Dom}(p_1)|, |\operatorname{Dom}(p_2)| \leq k+1$, then the notion of reaching positions can also be stated in terms of resolution: Spoiler can (regularly) reach $p_2$ from $p_1$ on $\Gamma$ if, and only if, there is a (regular) width-$(k+1)$ resolution derivation of $\gamma_{p_1}$ from $\Gamma \cup \{\gamma_{p_2}\}$, where $\gamma_p$ is the maximal clause falsified by $p$.

It is more difficult to establish a winning strategy for Duplicator, but we can benefit from the view of the width-$(k+1)$ game as existential $(k+2)$-pebble game [2] and the techniques developed for the existential pebble games in [7].

**Definition 13.** A *critical strategy* for Duplicator in the width-$(k+1)$ game on $\Gamma$ is a nonempty family $\mathcal{H}$ of partial assignments that do not falsify any clause from $\Gamma$ and a set of *critical positions* $\operatorname{crit}(\mathcal{H}) \subseteq \mathcal{H}$ such that:

- $p \in \operatorname{crit}(\mathcal{H}) \Rightarrow |\operatorname{Dom}(p)| = k+1$.
- If $p \in \mathcal{H}$ and $p' \subset p$, then $p' \in \mathcal{H}$.
- For every $p \in \mathcal{H} \setminus \operatorname{crit}(\mathcal{H})$, $|\operatorname{Dom}(p)| \leq k+1$, and every variable $Z \in \operatorname{Var}(\Gamma)$ there is a value $z \in \{0,1\}$ such that $p \cup \{Z \mapsto z\} \in \mathcal{H}$.

If $\operatorname{crit}(\mathcal{H}) = \emptyset$, then $\mathcal{H}$ is a *winning strategy.*

If there is a winning strategy $\mathcal{H}$ for Duplicator, then she can always provide a correct answer $z$ for a queried variable $Z$ without falsifying any clause from $\Gamma$. A critical strategy is nearly a winning strategy in the sense that Duplicator wins unless the game reaches a critical position. Duplicator may not have an appropriate answer in that situation, but she knows that Spoiler has stored a critical position (and nothing else, since $|\operatorname{Dom}(p)| = k+1$) and can use this information to flip to another critical strategy $\mathcal{H}'$ with $p \in \mathcal{H}'$. The following lemma enables us to construct a winning strategy out of a collection of critical strategies.

**Lemma 14.** *If $\mathcal{H}_1, \dots, \mathcal{H}_l$ are critical strategies on $\Gamma$ and for all $i \in [l]$ and all $p \in \operatorname{crit}(\mathcal{H}_i)$ there exists a $j \in [l]$ such that $p \in \mathcal{H}_j \setminus \operatorname{crit}(\mathcal{H}_j)$, then $\bigcup_{i\in[l]} \mathcal{H}_i$ is a winning strategy on $\Gamma$.* □

Every gadget $Q \subseteq \Gamma(G)$ we construct has a *boundary* $\operatorname{bd}(Q) \subseteq \operatorname{Var}(Q)$, that are the variables on which the gadget is connected to other gadgets. Furthermore, two gadgets $Q$ and $Q'$ are only connected by the clauses $\{X, \neg Y\}$ and $\{\neg X, Y\}$ (denoted $X \leftrightarrow Y$) for variables $X \in \operatorname{bd}(Q)$ and $Y \in \operatorname{bd}(Q')$. A *boundary function* of a strategy $\mathcal{H}$ on a gadget $Q$ is a function $\beta\colon \operatorname{bd}(Q) \to \{0,1\}$ such that $p(X) = \beta(X)$ for all $p \in \mathcal{H}$ and $X \in \operatorname{bd}(Q) \cap \operatorname{Dom}(p)$. We say that two strategies $\mathcal{G}$ and $\mathcal{H}$ on gadgets $Q^{\mathcal{G}}$ and $Q^{\mathcal{H}}$ are *connectable*, if they have boundary functions $\beta_{\mathcal{G}}$ and $\beta_{\mathcal{H}}$ and it holds that $\beta_{\mathcal{G}}(X) = \beta_{\mathcal{H}}(Y)$ for all $(X \leftrightarrow Y) \in \Gamma(G)$, $X \in \operatorname{bd}(Q^{\mathcal{G}})$, $Y \in \operatorname{bd}(Q^{\mathcal{H}})$.

**Lemma 15.** *Let $\mathcal{G}$ and $\mathcal{H}$ be two connectable critical strategies on gadgets $Q^{\mathcal{G}}$ and $Q^{\mathcal{H}}$. The composition $\mathcal{G} \uplus \mathcal{H} := \{g \cup h \mid g \in \mathcal{G}, h \in \mathcal{H}\}$ is a critical strategy on $Q^{\mathcal{G}} \cup Q^{\mathcal{H}}$ and their connecting clauses. Furthermore, $\mathcal{G} \uplus \mathcal{H}$ has critical positions* $\mathrm{crit}(\mathcal{G}) \cup \mathrm{crit}(\mathcal{H})$ *and the boundary function* $\beta_{\mathcal{G}} \cup \beta_{\mathcal{H}}$. □

We use the operator $\uplus$ to construct a critical strategy for $\Gamma(G)$ out of critical strategies on the gadgets. Then we show that the union of those global critical strategies is by Lemma 14 a winning strategy for Duplicator.

## 3.2. The Construction

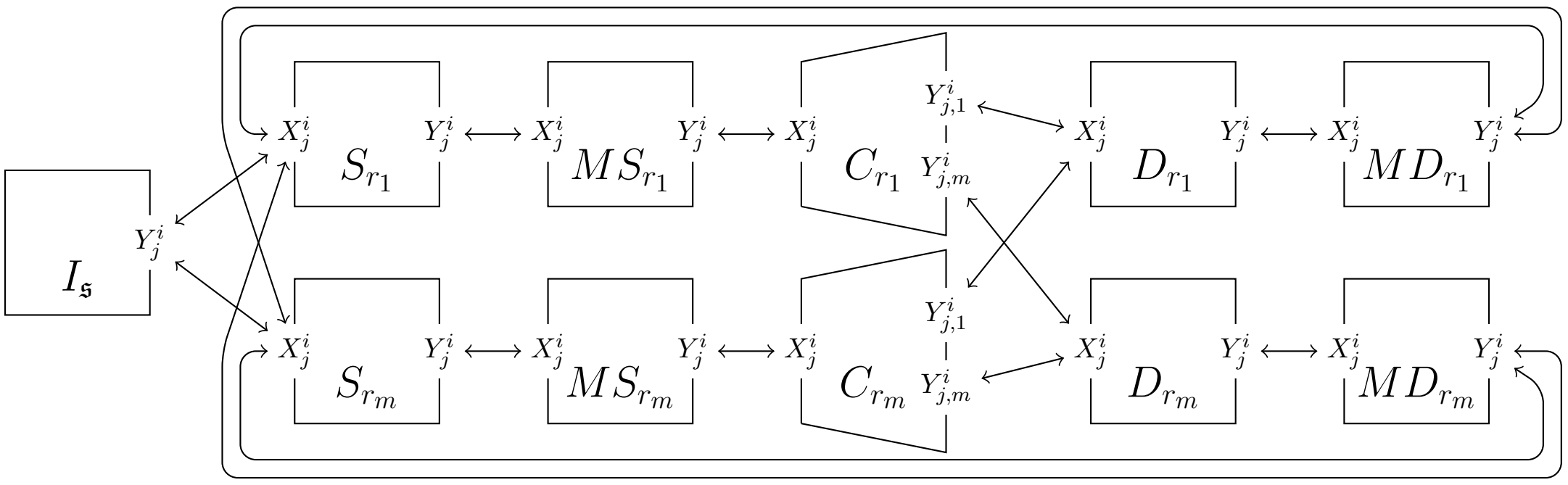

Figure 1: The 3-CNF formula $\Gamma(G)$.

In this paragraph we give an overview on the construction and the gadgets we use. Detailed descriptions of the gadgets and the strategies on them are given in the next section. We construct $\Gamma(G)$ as illustrated in Figure 1. The gadgets and their boundary variables are depicted as boxes and the arrows indicate the connection of the boundary variables. To encode the positions of the KAI-game we introduce Boolean variables $X^i_j$ for $i \in [k]$ and $j \in [n]$, which state "pebble $i$ is on node $j$". Every position $\mathfrak{p}$ is encoded by the partial assignment $\{X^i_{\mathfrak{p}(i)} \mapsto 1 \mid i \in [k]\}$, which will be denoted "position $\mathfrak{p}$ on $X$". A partial assignment of the variables $X^i_j$ is *invalid*, if there is at least on partition $l$ such that no variable $X^l_j$ is mapped to 1. The boundary of every gadget we construct consists of these variable blocks and we connect two blocks of variables $X^i_j$ and $Y^i_j$ by introducing clauses $X^i_j \leftrightarrow Y^i_j$ (for $i \in [k], j \in [n]$). If two blocks are connected in such a way, then Spoiler can regularly reach $\mathfrak{p}$ on $X$ from $\mathfrak{p}$ on $Y$ and vice versa. In order to do that, Spoiler stores $\mathfrak{p}$ on $X$ and then asks for $Y^1_{\mathfrak{p}(1)}$. Duplicator has to answer with 1 since otherwise this would falsify the clause $\{\neg X^1_{\mathfrak{p}(1)}, Y^1_{\mathfrak{p}(1)}\}$. Next, Spoiler deletes the assignment $X^1_{\mathfrak{p}(1)} \mapsto 1$ and asks for $Y^2_{\mathfrak{p}(2)}$. Once again, Duplicator has to answer with 1. Following that strategy Spoiler can regularly reach $\mathfrak{p}$ on $Y$ from $\mathfrak{p}$ on $X$. We want the players to move positions from left to right through the gadgets, that is they first store a position on the *input* boundary $X$ on the left side, then they play on the gadget and finally they reach a position on the *output* boundary $Y$ on the right side.

The *Initialization Gadget* $I_{\mathfrak{s}}$ is used to start the game. It has boundary variables $Y(I_{\mathfrak{s}})^i_j$ ($i \in [k]$, $j \in [n]$) and the feature that Spoiler can regularly reach $\mathfrak{s}$ on $Y(I_{\mathfrak{s}})$, the assignment that encodes the start position of the KAI-game.

For every rule $r$ there is a *Rule Gadget for Spoiler* $S_r$ with input boundary variables $X(S_r)^i_j$ and output boundary variables $Y(S_r)^i_j$. This gadget is used to modify the current KAI-game position according to rule $r$. If $r$ is applicable to $\mathfrak{p}$, then Spoiler can regularly reach $r(\mathfrak{p})$ on $Y(S_r)$ from $\mathfrak{p}$ on $X(S_r)$ and he does this whenever Player 1 applies rule $r$ to position $\mathfrak{p}$ in the KAI-game. Since Player 1 starts the KAI-game, the input $X(S_r)^i_j$ of every $S_r$ is connected to the output $Y(I_{\mathfrak{s}})^i_j$ of the Initialization Gadget. Hence, Spoiler can reach the start position on the input of every $S_r$. If a position $\mathfrak{p}$ is on the input variables of $S_r$ for a rule $r$ that is not applicable to $\mathfrak{p}$, then Duplicator has a strategy to avoid valid positions at the output of $S_r$, i.e. there exists a partition $l$ such that no variable $Y(S_r)^l_j$ is mapped to 1. We use this fact to force Spoiler to choose only applicable rules as it is the case for Player 1 in the KAI-game.

After every Rule Gadget $S_r$ there is a copy $MS_r$ of the *Switch* $M$ with input variables $X(M)^i_j$ and output variables $Y(M)^i_j$. Switches were already used before to prove lower bounds for model theoretic pebble games [7, 10, 16] and they are always the most involved part of the construction. This holds also for our Switch that bases on some kind of pigeonhole principle. From a valid position $\mathfrak{p}$ on $X(M)$ Spoiler can reach $\mathfrak{p}$ on $Y(M)$, but he cannot move invalid positions through the Switch. Duplicator's *impasse strategies* ensure that from an invalid partial assignment on the input variables (where no variable $X(M)^l_j$ is mapped to 1 for at least one $l$) Spoiler can only reach positions that map output variables to 0. Especially, moving $\mathfrak{p}$ through $S_r$ for a rule $r$ not applicable to $\mathfrak{p}$ leads to an invalid position on the output of $S_r$ and on the input of $MS_r$ and hence to an impasse. Another property of the Switch is that Spoiler has to reach a critical position inside the Switch in order to move a valid position from the input to the output and thus cannot store assignments outside of the Switch. Moreover, Spoiler cannot reach a position on the input from a position on the output. It follows that once Spoiler moves a position from left to right through the Switch he cannot move backwards and has no information about the variables outside of the Switch.

After every Switch there is a copy $C_r$ of the *Choice Gadget* $C$ that enables Duplicator to chose the next rule. This choice corresponds to the choice of Player 2 in the KAI-game. The Choice Gadget has one block of input variables $X(C)^i_j$ and for every rule $r_l$ a block of output variables $Y(C)^i_{j,l}$. First, if the current position $\mathfrak{p}$ on $X(C)$ is already a winning position for Player 1 ($\mathfrak{p}(i) = \theta$ for some $i \in [k]$), then Spoiler wins immediately. To ensure that we introduce clauses $\{\neg X(C)^i_\theta\}$ for every $i \in [k]$ in $C$. Second, Spoiler can regularly reach $\{Y(C)^i_{\mathfrak{p}(i),q} \mapsto 1 \mid i \in [k]\}$ from $\mathfrak{p}$ on $X(C)$ for some $q \in [m]$ of Duplicator's choice. Duplicator has for every rule $r_q$ a strategy to answer with $\mathfrak{p}$ on the input variables and on the $q$-th block of the output variables, and with 0 on all other output variables.

The $q$-th block of output variables of every Choice Gadget $C$ is connected to input variables of the corresponding *Rule Gadget for Duplicator* $D_{r_q}$. Analog to $S_r$ these gadgets have input variables $X(D_r)^i_j$ and output variables $Y(D_r)^i_j$, and Spoiler can regularly reach position $r(\mathfrak{p})$ on $Y(D_r)$ from $\mathfrak{p}$ on $X(D_r)$. If Duplicator has chosen a rule $r$ not applicable to the current position $\mathfrak{p}$, then Spoiler wins immediately from $\mathfrak{p}$ on $X(D_r)$. There are Switches also after the $D_r$ gadgets and the output variables of that Switches are connected to the input variables of the $S_r$ gadgets. Hence, Spoiler can move to the rule gadget $S_r$ that corresponds to Player 1's next choice. By playing the way described above, Spoiler can simulate a play of the KAI-game. If this play ends up with a winning position for Player 1, then Spoiler wins the game by falsifying some clause $\{\neg X(C)^i_\theta\}$. Duplicator's strategies ensure that this is the only way for Spoiler to win the game.

# 4. The Gadgets

For a partial assignment $p$ we let $\mathrm{cl}(p) := \{p' \mid p' \subseteq p\}$. It is easy to see that if $p$ is a satisfying total assignment of $\Gamma$, then $\mathrm{cl}(p)$ is a winning strategy for Duplicator in the width-$(k+1)$ game on $\Gamma$.

## 4.1. Rule Gadget for Spoiler

For every rule $r = (u, v, w, c, d)$ the Rule Gadget for Spoiler $S_r$ consists of variables $X(S_r)^i_j$ and $Y(S_r)^i_j$ for all $i \in [k]$ and $j \in [n]$ that are all boundary variables. There are the following clauses:

$$X(S_r)^c_u \to Y(S_r)^c_w, \tag{1}$$
$$X(S_r)^d_v \to Y(S_r)^d_v, \tag{2}$$
$$X(S_r)^i_j \to Y(S_r)^i_j, \qquad i \in [k] \setminus \{c, d\}, j \in [n] \setminus \{w\}. \tag{3}$$

**Lemma 16** (Spoiler's strategy on $S_r$)**.** *Spoiler can regularly reach $\mathfrak{p}$ on $Y(S_r)$ from $\mathfrak{p}$ on $X(S_r)$ for every position $\mathfrak{p}$ and every rule $r$ applicable to $\mathfrak{p}$.*

*Proof.* By definition, the gadget contains the clauses $X(S_r)^i_{\mathfrak{p}(i)} \to Y(S_r)^i_{r(\mathfrak{p})(i)}$ for $i \in [k]$. Thus, starting from position $\{X(S_r)^i_{\mathfrak{p}(i)} \mapsto 1 \mid i \in [k]\}$ Spoiler can ask for $Y(S_r)^1_{\mathfrak{p}(1)}$ and Duplicator has to answer with $Y(S_r)^1_{\mathfrak{p}(1)} \mapsto 1$. Now, Spoiler deletes $X(S_r)^1_{\mathfrak{p}(1)}$ and asks for $Y(S_r)^2_{\mathfrak{p}(2)}$. Once more, Duplicator has to answer with 1. Following that strategy, Spoiler can regularly reach $\{Y(S_r)^i_{r(\mathfrak{p})(i)} \mapsto 1 \mid i \in [k]\}$. ☐

The next lemma states that Duplicator does not lose when Spoiler moves through the gadget. Furthermore, if Spoiler has chosen a Rule Gadget $S_r$ and the rule $r$ is not applicable to the current position $\mathfrak{p}$ (hence $T_r(\mathfrak{p}) \neq \emptyset$), then Duplicator has a strategy that avoids valid positions at the output.

**Lemma 17** (Duplicator's strategies on $S_r$)**.** *For every position $\mathfrak{p}$ Duplicator has a winning strategy $\mathcal{R}_\mathfrak{p}$ with boundary function $\beta_\mathfrak{p}(X(S_r)^i_j) = 1$, iff $j = \mathfrak{p}(i)$; and $\beta_\mathfrak{p}(Y(S_r)^i_j) = 1$, iff $i \notin T_r(\mathfrak{p})$ and $j = r(\mathfrak{p})(i)$. Furthermore, she has a winning strategy $\mathcal{R}_0$ with boundary function $\beta_0(X^i_j) = \beta_0(Y^i_j) = 0$ for all $i \in [k]$ and $j \in [n]$.*

*Proof.* Let $\mathcal{R}_\mathfrak{p} := \mathrm{cl}(\beta_\mathfrak{p})$ and $\mathcal{R}_0 := \mathrm{cl}(\beta_0)$, where $\beta_\mathfrak{p}$ and $\beta_0$ are the boundary functions defined in the above lemma. Since $\beta_\mathfrak{p}$ and $\beta_0$ define total assignments that satisfy all clauses from the gadget, $\mathcal{R}_\mathfrak{p}$ and $\mathcal{R}_0$ are winning strategies on $S_r$. ☐

## 4.2. Rule Gadget for Duplicator

For every rule $r = (u, v, w, c, d)$ the Rule Gadget for Duplicator $D_r$ consists of boundary variables $X(D_r)^i_j$ and $Y(D_r)^i_j$ for all $i \in [k]$ and $j \in [n]$ and the following clauses:

$$X(D_r)^c_u \to Y(D_r)^c_w, \tag{4}$$
$$X(D_r)^d_v \to Y(D_r)^d_v, \tag{5}$$
$$X(D_r)^i_j \to Y(D_r)^i_j, \qquad i \in [k] \setminus \{c, d\}; j \in [n] \setminus \{w\}, \tag{6}$$
$$\neg X(D_r)^c_j, \qquad j \neq u, \tag{7}$$
$$\neg X(D_r)^d_j, \qquad j \neq v, \tag{8}$$
$$\neg X(D_r)^i_w, \qquad i \in [k] \setminus \{c, d\}. \tag{9}$$

As for the $S_r$ gadget, Spoiler can move a valid position through the gadget while applying the rule. If Duplicator has chosen a Rule Gadget for a rule $r$ not applicable to $\mathfrak{p}$, then she is penalized by losing immediately.

**Lemma 18** (Spoiler's strategy on $D_r$)**.** *Spoiler can regularly reach $\mathfrak{p}$ on $Y(D_r)$ from $\mathfrak{p}$ on $X(D_r)$ for every position $\mathfrak{p}$ and every rule $r$ applicable to $\mathfrak{p}$. Furthermore, if $r$ is not applicable to $\mathfrak{p}$, then Spoiler wins from position $\mathfrak{p}$ on $X(D_r)$.*

*Proof.* If $r$ is applicable to $\mathfrak{p}$, then there are clauses $X(S_r)^i_{\mathfrak{p}(i)} \to Y(S_r)^i_{r(\mathfrak{p})(i)}$ for $i \in [k]$. Spoiler can regularly reach $\{Y(D_r)^i_{r(\mathfrak{p})(i)} \mapsto 1 \mid i \in [k]\}$ from $\{X(D_r)^i_{\mathfrak{p}(i)} \mapsto 1 \mid i \in [k]\}$ analog to Lemma 16. If $r$ is not applicable to $\mathfrak{p}$, then $\{X(D_r)^i_{\mathfrak{p}(i)} \mapsto 1 \mid i \in [k]\}$ falsifies some clause from (7)-(9). ☐

The next lemma states that Duplicator does not lose the game if the rule is applicable to the current position or if all variables are mapped to 0.

**Lemma 19** (Duplicator's strategies on $D_r$). *If $r$ is applicable to $\mathfrak{p}$, then Duplicator has a winning strategy $\mathcal{R}_\mathfrak{p}$ on $D_r$ with boundary function $\beta_\mathfrak{p}(X(D_r)^i_j) = 1$, iff $j = \mathfrak{p}(i)$; and $\beta_\mathfrak{p}(Y(D_r)^i_j) = 1$, iff $j = r(\mathfrak{p})(i)$. Furthermore, she has a winning strategy $\mathcal{R}_0$ with boundary function $\beta_0(X(D_r)^i_j) = \beta_0(Y(D_r)^i_j) = 0$ for all $i \in [k]$ and $j \in [n]$.*

*Proof.* Analog to Lemma 17, let $\mathcal{R}_\mathfrak{p} := \mathrm{cl}(\beta_\mathfrak{p})$ and $\mathcal{R}_0 := \mathrm{cl}(\beta_0)$, where $\beta_\mathfrak{p}$ and $\beta_0$ are the boundary functions defined in the above lemma. ☐

## 4.3. The Switch

The Switch $M$ contains input variables $X(M)^i_j$, output variables $Y(M)^i_j$ and additional variables inside. The clauses of the Switch are given below for all $i, i', l \in [k]$, $j, j' \in [n]$ and $c, c' \in \{1, 2, 3, 4\}$.

$$X(M)^i_j \to A0^i_j \vee A1^i_j \tag{10}$$

$$A0^i_j \to A^{i,1}_j \vee A^{i,2}_j \tag{11}$$

$$A1^i_j \to A^{i,3}_j \vee A^{i,4}_j \tag{12}$$

$$A^{i,c}_j \to A^{i,c}_{j,1} \vee A^{i,c}_{j,\geq 2} \tag{13}$$

$$A^{i,c}_{j,\geq l} \to A^{i,c}_{j,l} \vee A^{i,c}_{j,\geq l+1} \qquad 2 \leq l \leq k-2 \tag{14}$$

$$A^{i,c}_{j,\geq k-1} \to A^{i,c}_{j,k-1} \vee A^{i,c}_{j,k} \tag{15}$$

$$\neg\left(A^{i,c}_{j,l} \wedge A^{i',c'}_{j',l}\right) \qquad i \neq i' \tag{16}$$

$$A^{i,c}_{j,l} \to B_l \tag{17}$$

$$B_1 \wedge B_{\geq 2} \to B \tag{18}$$

$$B_l \wedge B_{\geq l+1} \to B_{\geq l} \qquad 2 \leq l \leq k-2 \tag{19}$$

$$B_{k-1} \wedge B_k \to B_{\geq k-1} \tag{20}$$

$$A^{i,c}_{j,l} \wedge B \to Y(M)^i_j \tag{21}$$

The essence of the Switch can be described by a kind of pigeon hole principle. There are $k$ holes and $kn$ groups of four pigeons each. Every group of four pigeons corresponds to one of the $kn$ variables $X(M)^i_j$. The four pigeons in the pigeon group $X(M)^i_j$ correspond to the four variables $A^{i,1}_j$, $A^{i,2}_j$, $A^{i,3}_j$ and $A^{i,4}_j$. The variables $A^{i,c}_j$ determine whether the corresponding pigeon is *arriving*. Variable $X(M)^i_j$ says that one pigeon $A^{i,c}_j$ of the pigeon group is arriving (stated by the clauses (10), (11) and (12)). Thus, a partial assignment $\{X(M)^i_{\mathfrak{p}(i)} \mapsto 1 \mid i \in [k]\}$ forces $k$ pigeons to arrive. The variables $A^{i,c}_{j,l}$ say "pigeon $A^{i,c}_j$ sits in hole $l$". It is ensured by the clauses (13), (14) and (15) that if $A^{i,c}_j$ is arriving, then it will sit in some hole. The clauses (16) state that in every hole there is at most one pigeon.

The intended meaning of the variable $B_l$ is "hole $l$ is occupied" and it is ensured by the clauses (17) that this variable is true, if some pigeon actually sits in hole $l$. The variable $B$ states "all holes are occupied" and it is guaranteed by the clauses (18), (19) and (20) that $B$ is true, if all $B_l$ are true. The clauses (21) state that if all holes are occupied and pigeon $A_j^{i,c}$ (from the pigeon group $X(M)_j^i$) sits in some hole, then $Y(M)_j^i$ has to be true. Moving a position $\mathfrak{p}$ through the Switch proceeds, roughly, in the following way. At the beginning the partial assignment $\mathfrak{p}$ is on the input $X(M)$. There sits no pigeon in any hole and Duplicator plays according to a critical *input strategy* that maps all output variables to 0. In order to reach $\mathfrak{p}$ on the output, Spoiler has to bring all pigeons into the pigeon house. He can force Duplicator to decide which pigeon from the corresponding pigeon group is arriving and then he forces Duplicator to specify a mapping from the $k$ arriving pigeons to the $k$ holes. Unless the $k$-th pigeon is arriving, Duplicator maintains $B \mapsto 0$ and thus he can maintain $Y(M)_j^i \mapsto 0$ without falsifying any clause. As soon as every pigeon is arriving the game reaches a critical position. At this point Duplicator flips to an *output strategy* with $B \mapsto 1$ and $Y(M)_{\mathfrak{p}(i)}^i \mapsto 1$. On the other hand, he flips all input variables $X(M)_j^i$ to 0 and hence prevents Spoiler from reaching any position at the input.

If there is an invalid position at the input vertices, then at most $k-1$ variables $X(M)_j^i$ are mapped to 1. Thus, Spoiler can force only $k-1$ pigeons to arrive. Since in that case at most $k-1$ holes are occupied, Duplicator use an *impasse strategy* to maintain $B \mapsto 0$ and $Y(M)_j^i \mapsto 0$ without contradicting any clause. Therefore, Spoiler cannot move invalid positions through the Switch. The next two lemmas state Spoiler's and Duplicator's strategies formally, the proofs are deferred to the appendix.

**Lemma 20** (Spoiler's strategy on $M$). *Spoiler can regularly reach $\mathfrak{p}$ on $Y(M)$ from $\mathfrak{p}$ on $X(M)$.*

**Lemma 21** (Duplicator's strategies on $M$). *For every position $\mathfrak{p}$ and every nonempty $T \subseteq [k]$, there are strategies $\mathcal{S}_{\mathfrak{p},T}^{imp}$, $\mathcal{S}_{\mathfrak{p}}^{out}$ and $\mathcal{S}_{\mathfrak{p}}^{in}$ for Duplicator satisfying the following conditions.*

*(i) The impasse strategy $\mathcal{S}_{\mathfrak{p},T}^{imp}$ is a winning strategy with boundary function $\beta(X(M)_j^i) = 1$, iff $i \notin T$ and $j = \mathfrak{p}(i)$; and $\beta(Y(M)_j^i) = 0$, for all $i \in [k], j \in [n]$.*

*(ii) The output strategy $\mathcal{S}_{\mathfrak{p}}^{out}$ is a winning strategy with boundary function $\beta(X(M)_j^i) = 0$, for all $i \in [k]$, $j \in [n]$; and $\beta(Y(M)_j^i) = 1$, iff $j = \mathfrak{p}(i)$.*

*(iii) The input strategy $\mathcal{S}_{\mathfrak{p}}^{in}$ is a critical strategy with $\mathrm{crit}(\mathcal{S}_{\mathfrak{p}}^{in}) \subseteq \mathcal{S}_{\mathfrak{p}}^{out}$ and boundary function $\beta(X(M)_j^i) = 1$, iff $j = \mathfrak{p}(i)$; and $\beta(Y(M)_j^i) = 0$, for all $i \in [k], j \in [n]$.*

## 4.4. The Initialization Gadget

For a start position $\mathfrak{s}$ the Initialization Gadget $I_{\mathfrak{s}}$ consists of two Switches $M_1$ and $M_2$, start variables $S_1$ and $S_2$, and boundary variables $Y(I_{\mathfrak{s}})_j^i$ for all $i \in [k]$ and $j \in [n]$. There are the following clauses in addition to the ones of $M_1$ and $M_2$:

$$S_1 \vee S_2 \tag{22}$$

$$S_1 \to X(M_c)_{\mathfrak{s}(i)}^i, \text{ for all } i \in [k],\ c \in \{1, 2\} \tag{23}$$

$$Y(M_c)_{\mathfrak{s}(i)}^i \to Y(I_{\mathfrak{s}})_{\mathfrak{s}(i)}^i, \text{ for all } i \in [k],\ c \in \{1, 2\} \tag{24}$$

**Lemma 22** (Spoiler's strategy on $I_{\mathfrak{s}}$). *Spoiler can regularly reach $\mathfrak{s}$ on $Y(I_{\mathfrak{s}})$.*

*Proof.* First, Spoiler pebbles $S_1$ and $S_2$. Because of clause $S_1 \vee S_2$, Duplicator has to answer 1 for $S_1$ or $S_2$. Depending on Duplicator's choice, Spoiler can either reach $\mathfrak{s}$ on $X(M_1)$ or $\mathfrak{s}$ on $X(M_2)$ owing to clauses (23). By applying Lemma 20 Spoiler can reach $\mathfrak{s}$ on $Y(M_1)$ ($\mathfrak{s}$ on $Y(M_2)$) and thus he can reach $\mathfrak{s}$ on $Y(I_{\mathfrak{s}})$ using clauses (24). □

We can combine the strategies from Lemma 21 on the switches $M_1$ and $M_2$ to obtain strategies for Duplicator on $I_{\mathfrak{s}}$. The winning strategy $\mathcal{I}^{\text{init}}$ says that Duplicator does not lose when Spoiler reaches $\mathfrak{s}$ on $Y(I_{\mathfrak{s}})$. Duplicator uses the critical strategies $\mathcal{I}_{\mathfrak{p}}^{\text{init}}$ and $\mathcal{I}_0^{\text{init}}$ if other positions than the start position occur at the output of $I_{\mathfrak{s}}$ during the course of the game.

**Lemma 23** (Duplicator's strategies on $I_{\mathfrak{s}}$)**.** *There are strategies $\mathcal{I}^{init}$, $\mathcal{I}^{init}_{\mathfrak{p}}$ and $\mathcal{I}^{init}_0$ for Duplicator with the following properties.*

*(i) $\mathcal{I}^{init}$ is a winning strategy with boundary function $\beta(Y(I_{\mathfrak{s}})^i_j) = 1$, iff $j = \mathfrak{s}(i)$.*

*(ii) $\mathcal{I}^{init}_{\mathfrak{p}}$ is a critical strategy with $\mathrm{crit}(\mathcal{I}^{init}_{\mathfrak{p}}) \subseteq \mathcal{I}^{init}$ and boundary function $\beta_{\mathfrak{p}}(Y(I_{\mathfrak{s}})^i_j) = 1$, iff $j = \mathfrak{p}(i)$.*

*(iii) $\mathcal{I}^{init}_0$ is a critical strategy with $\mathrm{crit}(\mathcal{I}^{init}_0) \subseteq \mathcal{I}^{init}$ and boundary function $\beta_0(Y(I_{\mathfrak{s}})^i_j) = 0$ for all boundary variables $Y(I_{\mathfrak{s}})^i_j$.*

*Proof.* Recall the strategies $\mathcal{S}^{\mathrm{out}}_{\mathfrak{s}}$ and $\mathcal{S}^{\mathrm{in}}_{\mathfrak{s}}$ from Lemma 21.

$$\begin{aligned}
\mathcal{I}_1 &:= \mathcal{S}^{\mathrm{in}}_{\mathfrak{s}}(M_1) \uplus \mathcal{S}^{\mathrm{out}}_{\mathfrak{s}}(M_2) \uplus \mathrm{cl}\big(\{S_1 \mapsto 1, S_2 \mapsto 0\} \cup \\
&\qquad \{Y^i_{\mathfrak{s}(i)} \mapsto 1 \mid i \in [k]\} \cup \\
&\qquad \{Y^i_j \mapsto 0 \mid i \in [k], j \in [n], j \neq \mathfrak{s}(i)\}\big) \\
\mathcal{I}_2 &:= \mathcal{S}^{\mathrm{out}}_{\mathfrak{s}}(M_1) \uplus \mathcal{S}^{\mathrm{in}}_{\mathfrak{s}}(M_2) \uplus \mathrm{cl}\big(\{S_1 \mapsto 0, S_2 \mapsto 1\} \cup \\
&\qquad \{Y^i_{\mathfrak{s}(i)} \mapsto 1 \mid i \in [k]\} \cup \\
&\qquad \{Y^i_j \mapsto 0 \mid i \in [k], j \in [n], j \neq \mathfrak{s}(i)\}\big) \\
\mathcal{I}^{\mathrm{init}} &:= \mathcal{I}_1 \cup \mathcal{I}_2
\end{aligned}$$

By Lemma 15, $\mathcal{I}_1$ and $\mathcal{I}_2$ are critical strategies with $\mathrm{crit}(\mathcal{I}_1) = \mathrm{crit}(\mathcal{S}^{\mathrm{in}}_{\mathfrak{s}}(M_1))$ and $\mathrm{crit}(\mathcal{I}_2) = \mathrm{crit}(\mathcal{S}^{\mathrm{in}}_{\mathfrak{s}}(M_2))$. From

$$\begin{aligned}
\mathrm{crit}(\mathcal{I}_1) = \mathrm{crit}(\mathcal{S}^{\mathrm{in}}_{\mathfrak{s}}(M_1)) \subseteq \mathcal{S}^{\mathrm{out}}_{\mathfrak{s}}(M_2) \subseteq \mathcal{I}_2 \setminus \mathrm{crit}(\mathcal{I}_2) \text{ and} \\
\mathrm{crit}(\mathcal{I}_2) = \mathrm{crit}(\mathcal{S}^{\mathrm{in}}_{\mathfrak{s}}(M_2)) \subseteq \mathcal{S}^{\mathrm{out}}_{\mathfrak{s}}(M_1) \subseteq \mathcal{I}_1 \setminus \mathrm{crit}(\mathcal{I}_1)
\end{aligned}$$

it follows that $\mathcal{I}^{\mathrm{init}}$ is a winning strategy by Lemma 14. This proves (i), to establish (ii) and (iii) let

$$\begin{aligned}
\mathcal{I}^{\mathrm{init}}_{\mathfrak{p}} &:= \mathcal{S}^{\mathrm{in}}_{\mathfrak{s}}(M_1) \uplus \mathcal{S}^{\mathrm{in}}_{\mathfrak{s}}(M_2) \uplus \mathrm{cl}\big(\{S_1 \mapsto 1, S_2 \mapsto 1\} \cup \\
&\qquad \{Y^i_{\mathfrak{p}(i)} \mapsto 1 \mid i \in [k]\} \cup \\
&\qquad \{Y^i_j \mapsto 0 \mid i \in [k], j \in [n], j \neq \mathfrak{p}(i)\}\big) \text{ and} \\
\mathcal{I}^{\mathrm{init}}_0 &:= \mathcal{S}^{\mathrm{in}}_{\mathfrak{s}}(M_1) \uplus \mathcal{S}^{\mathrm{in}}_{\mathfrak{s}}(M_2) \uplus \mathrm{cl}\big(\{S_1 \mapsto 1, S_2 \mapsto 1\} \cup \\
&\qquad \{Y^i_j \mapsto 0 \mid i \in [k], j \in [n]\}\big).
\end{aligned}$$

Lemma 15 tells us that $\mathcal{I}^{\mathrm{init}}_{\mathfrak{p}}$ and $\mathcal{I}^{\mathrm{init}}_0$ are critical strategies with $\mathrm{crit}(\mathcal{I}^{\mathrm{init}}_0) = \mathrm{crit}(\mathcal{I}^{\mathrm{init}}_{\mathfrak{p}}) = \mathrm{crit}(\mathcal{S}^{\mathrm{in}}_{\mathfrak{s}}(M_1)) \cup \mathrm{crit}(\mathcal{S}^{\mathrm{in}}_{\mathfrak{s}}(M_2))$. Therefore, $\mathrm{crit}(\mathcal{I}^{\mathrm{init}}_{\mathfrak{p}}) \subseteq \mathcal{I}^{\mathrm{init}}$ and $\mathrm{crit}(\mathcal{I}^{\mathrm{init}}_0) \subseteq \mathcal{I}^{\mathrm{init}}$. ☐

## 4.5. The Choice Gadget

The Choice Gadget $C$ contains input variables $X(C)^i_j$ for $i \in [k]$, $j \in [n]$ and output variables $Y(C)^i_{j,q}$ for all $i \in [k]$, $j \in [n]$ and $q \in [m]$ as boundary. Furthermore there are inner variables $E^i_{j,\geq q}$ for $i \in [k]$, $j \in [n]$ and $2 \leq q \leq m-1$. The clauses are given below.

$$\begin{aligned}
&\neg X(C)^i_\theta && i \in [k] && (25) \\
&X(C)^i_j \rightarrow Y(C)^i_{j,1} \vee E^i_{j,\geq 2} && && (26) \\
&E^i_{j,\geq q} \rightarrow Y(C)^i_{j,q} \vee E^i_{j,\geq q+1} && 2 \leq q \leq m-2 && (27) \\
&E^i_{j,\geq m-1} \rightarrow Y(C)^i_{j,m-1} \vee Y(C)^i_{j,m} && && (28) \\
&\neg(Y(C)^i_{j,q} \wedge Y(C)^i_{j,q'}) && q \neq q' && (29)
\end{aligned}$$

**Lemma 24** (Spoiler's strategy on $C$)**.** *Spoiler can regularly reach $\{Y(C)^i_{\mathfrak{p}(i),q} \mapsto 1 \mid i \in [k]\}$ from $\{X(C)^i_{\mathfrak{p}(i)} \mapsto 1 \mid i \in [k]\}$ for some $q \in [m]$ of Duplicator's choice.*[1] *Moreover, if $\mathfrak{p}$ is a winning position for Player 1, then Spoiler wins immediately.*

*Proof.* Starting from $\{X(C)^i_{\mathfrak{p}(i)} \mapsto 1 \mid i \in [k]\}$ Spoiler picks up the two remaining pebbles and asks for $Y(C)^1_{\mathfrak{p}(1),1}$ and $E^1_{\mathfrak{p}(1),\geq 2}$. Owing to clause (26) Duplicator has to answer 1 for one of the two. If Duplicator does not answer with $Y(C)^1_{\mathfrak{p}(1),1} \mapsto 1$, then Spoiler moves the pebbles from $X(C)^1_{\mathfrak{p}(1)}$ and $Y(C)^1_{\mathfrak{p}(1),1}$ to $Y(C)^1_{\mathfrak{p}(1),2}$ and $E^1_{\mathfrak{p}(1),\geq 3}$. Because of clause (27) Duplicator has to answer with $Y(C)^1_{\mathfrak{p}(1),2} \mapsto 1$ or $E^1_{\mathfrak{p}(1),\geq 3} \mapsto 1$. Using that strategy Spoiler can reach $\{Y(C)^1_{\mathfrak{p}(1),q} \mapsto 1\} \cup \{X(C)^i_{\mathfrak{p}(i)} \mapsto 1 \mid 2 \leq i \leq k\}$ for some $q \in [m]$ of Duplicator's choice. In the next step Spoiler applies the same technique to the other partitions. If Duplicator chooses a $q' \neq q$ in an other partition, then she loses immediately owing to clause (29). Thus, Spoiler can reach $\{Y(C)^i_{\mathfrak{p}(i),q} \mapsto 1 \mid i \in [k]\}$. Since he has not pebbled a variable twice, this strategy is regular. In addition, if $\mathfrak{p}$ is a winning position for Spoiler, then $\{X(C)^i_{\mathfrak{p}(i)} \mapsto 1 \mid i \in [k]\}$ clearly falsifies some clause from (25). ☐

**Lemma 25** (Duplicator's strategies on $C$)**.** *For every position $\mathfrak{p}\colon [k] \to [n] \setminus \{\theta\}$ and every $q \in [m]$ there is a winning strategy $\mathcal{C}^q_{\mathfrak{p}}$ for Duplicator with boundary function $\beta^q_{\mathfrak{p}}(X(C)^i_j) = 1$, iff $j = \mathfrak{p}(i)$; and $\beta^q_{\mathfrak{p}}(Y(C)^i_{j,l}) = 1$, iff $j = \mathfrak{p}(i)$ and $l = q$. Furthermore, there is a winning strategy $\mathcal{C}_0$ with boundary function $\beta_0$ mapping all boundary variables to 0.*

*Proof.* Let $C^q_{\mathfrak{p}}$ be the total assignment consisting of $\beta^q_{\mathfrak{p}}$ together with

$$E^i_{\mathfrak{p}(i),\geq l} \mapsto \begin{cases} 1, \text{ if } j = \mathfrak{p}(i) \text{ and } l \leq q, \\ 0, \text{ else,} \end{cases}$$

and $C_0$ be the assignment that maps every variable in the gadget to 0. Since the assignments $C^q_{\mathfrak{p}}$ and $C_0$ falsify no clause, the strategies $\mathcal{C}^q_{\mathfrak{p}} := \mathrm{cl}(C^q_{\mathfrak{p}})$ and $\mathcal{C}_0 := \mathrm{cl}(C_0)$ are winning strategies with the desired boundary function. ☐

# 5. The Reduction

**Lemma 26** (Spoiler's global strategy)**.** *If Player 1 has a winning strategy in the (acyclic) $k$-pebble KAI-game on $G$, then Spoiler has a winning strategy in the (regular) width-$(k+1)$ game on $\Gamma(G)$.*

*Proof.* Assume that Player 1 has a winning strategy in the $k$-pebble KAI-game on $G$. We have to show that Spoiler can reach a position that falsifies a clause. First, Spoiler can reach $\mathfrak{s}$ on $Y(I_{\mathfrak{s}})$ via the Initialization Gadget. Let $r$ be the rule applicable to $\mathfrak{s}$ Player 1 chooses first in his strategy and $\mathfrak{p}_1 := r(\mathfrak{s})$. Spoiler can reach $\mathfrak{s}$ on $X(S_r)$ by the connection of the boundary. He can move through the Rule Gadget to $\mathfrak{p}_1$ on $Y(S_r)$ and hence to $\mathfrak{p}_1$ on $X(MS_r)$ since the boundary variables are connected. In the next step he moves through the Switch and reaches $\mathfrak{p}_1$ on $Y(MS_r)$ and then $\mathfrak{p}_1$ on $X(C_r)$. If position $\mathfrak{p}_1$ is a winning position for Player 1 in the KAI-game (that is, one pebble is on node $\theta$), then Spoiler wins immediately. Thus, assume that $\mathfrak{p}_1$ is no winning position and Player 2 chooses a rule $r$ in the KAI-game. At this point Spoiler forces Duplicator to choose a $q \in [m]$ such that he can reach $\{Y(C_r)^i_{\mathfrak{p}_1(i),q} \mapsto 1 \mid i \in [k]\}$ and hence $\mathfrak{p}_1$ on $X(D_{r_q})$. If Duplicator has chosen a $q \in [m]$ such that $r_q$ is not applicable to $\mathfrak{p}_1$, then Spoiler wins immediately, especially he wins if there is no rule applicable to $\mathfrak{p}_1$ and Player 2 is unable to move. Thus, let $r_q$ be applicable to $\mathfrak{p}_1$ and $\mathfrak{p}_2 := r_q(\mathfrak{p}_1)$. Spoiler moves through the Rule Gadget, reaches $\mathfrak{p}_2$ on $Y(D_{r_q})$ and then $\mathfrak{p}_2$ on $X(MD_{r_q})$. Now he moves through the Switch to $\mathfrak{p}_2$ on $Y(MD_{r_q})$. Via the connection of the output variables of the Switch $MD_{r_q}$ to the input variables of the Rule Gadgets $S_r$, Spoiler chooses a rule $r$ that is applicable to $\mathfrak{p}_2$ and moves to $\mathfrak{p}_2$ on $X(S_r)$. The choice of the rule corresponds to the choice of Player 1 in his winning strategy. In the sequel, Spoiler applies that rule by moving through the Rule Gadget and so on. Simulating the strategy of Player 1 in this way, Spoiler can reach a position on the input variables of some Choice Gadget that encodes a winning position for Player 1 and thus falsifies a clause $\{\neg X(C)^i_\theta\}$.

[1] This statement in terms of resolution: There is a regular width-$(k+1)$ resolution derivation of $\{\neg X(C)^i_{\mathfrak{p}(i)} \mid i \in [k]\}$ from $\Gamma \cup \{\{\neg Y(C)^i_{\mathfrak{p}(i),q} \mid i \in [k]\} \mid q \in \{m\}\}$.

If $G$ is acyclic, then no rule can be applied twice. Thus, following that strategy above Spoiler does not play twice on one gadget. Since all partial strategies on the gadgets are regular this gives rise to a global regular strategy for Spoiler. □

**Lemma 27** (Duplicator's global strategy). *If Player 2 has a winning strategy in the $k$-pebble KAI-game on $G$, then Duplicator has a winning strategy in the width-$(k+1)$ game on $\Gamma(G)$.*

*Proof.* Let $\mathcal{K} = (\mathcal{K}_1, \mathcal{K}_2, \kappa)$ be a winning strategy for Player 2 in the $k$-pebble KAI-game on $G$. We construct a winning strategy $\mathcal{H}$ for Duplicator in the width-$(k+1)$ game on $\Gamma(G)$. First, we define auxiliary critical strategies $\mathcal{H}^1_\mathfrak{p}$, $\mathcal{H}^2_\mathfrak{p}$ and $\mathcal{H}^{\text{init}}$. In order to do this we combine Duplicator's strategies on the gadgets using the $\uplus$-operator and write $\mathcal{A}\langle B\rangle$ to pinpoint strategy $\mathcal{A}$ on gadget $B$. For all $\mathfrak{p} \in \mathcal{K}_1$ let

$$\begin{aligned}\mathcal{H}^1_\mathfrak{p} := \mathcal{I}^{\text{init}}_\mathfrak{p} \uplus \biguplus_r \big(\mathcal{C}_0\langle C_r\rangle \uplus \mathcal{R}_0\langle D_r\rangle \uplus \mathcal{S}^{\text{out}}_\mathfrak{p}\langle MD_r\rangle\big) \uplus \\ \biguplus_{r\in\text{appl}(\mathfrak{p})} \big(\mathcal{R}_\mathfrak{p}\langle S_r\rangle \uplus \mathcal{S}^{\text{in}}_{r(\mathfrak{p})}\langle MS_r\rangle\big) \uplus \\ \biguplus_{r\notin\text{appl}(\mathfrak{p})} \big(\mathcal{R}_\mathfrak{p}\langle S_r\rangle \uplus \mathcal{S}^{\text{imp}}_{r(\mathfrak{p}),T_r(\mathfrak{p})}\langle MS_r\rangle\big).\end{aligned}$$

The initialization strategy $\mathcal{H}^{\text{init}}$ differs from $\mathcal{H}^1_\mathfrak{s}$ only in the choice of the strategy on the Initialization Gadget: It contains the winning strategy $\mathcal{I}^{\text{init}}$ instead of the critical strategy $\mathcal{I}^{\text{init}}_\mathfrak{s}$. For all $\mathfrak{p} \in \mathcal{K}_2$ let

$$\begin{aligned}\mathcal{H}^2_\mathfrak{p} := \mathcal{I}^{\text{init}}_0 \uplus \biguplus_r \big(\mathcal{C}^{\kappa(\mathfrak{p})}_\mathfrak{p}\langle C_r\rangle \uplus \mathcal{R}_0\langle S_r\rangle \uplus \mathcal{S}^{\text{out}}_\mathfrak{p}\langle MS_r\rangle\big) \uplus \\ \biguplus_{r\neq r_{\kappa(\mathfrak{p})}} \big(\mathcal{R}_0\langle D_r\rangle \uplus \mathcal{S}^{\text{imp}}_{\mathfrak{p},[k]}\langle MD_r\rangle\big) \uplus \\ \big(\mathcal{R}_\mathfrak{p}\langle D_{r_{\kappa(\mathfrak{p})}}\rangle \uplus \mathcal{S}^{\text{in}}_{r_{\kappa(\mathfrak{p})}(\mathfrak{p})}\langle MD_{r_{\kappa(\mathfrak{p})}}\rangle\big).\end{aligned}$$

Note that all these strategies are by Lemma 15 global critical strategies. The strategies above enable Duplicator to simulate the moves of the KAI-game. Playing within strategy $\mathcal{H}^{\text{init}}$ means that the KAI-game has just started, position $\mathfrak{s}$ is on the board and it is Player 1's turn. Duplicator uses the strategy $\mathcal{H}^1_\mathfrak{p}$ ($\mathcal{H}^2_\mathfrak{p}$) to express that the current position in the KAI-game is $\mathfrak{p}$ and it is Player 1's (Player 2's) turn. If Spoiler reaches a critical position on the switches, then Duplicator flips the strategies in the same way as the positions in the KAI-game change. Let us now define the winning strategy $\mathcal{H}$ formally: $\mathcal{H} := \mathcal{H}^{\text{init}} \cup \bigcup_{\mathfrak{p}\in\mathcal{K}_1} \mathcal{H}^1_\mathfrak{p} \cup \bigcup_{\mathfrak{p}\in\mathcal{K}_2} \mathcal{H}^2_\mathfrak{p}$. To verify that $\mathcal{H}$ is indeed a winning strategy it remains to show, by Lemma 14, that every critical position in one auxiliary strategy is contained as non-critical position in some other auxiliary strategy. For a strategy $\mathcal{A}$ let $\hat{\mathcal{A}} := \mathcal{A} \setminus \text{crit}(\mathcal{A})$. We get the inclusions below by Lemma 21 and Lemma 23 for all $\mathfrak{p}_1 \in \mathcal{K}_1$ and $\mathfrak{p}_2 \in \mathcal{K}_2$.

$$\text{crit}(\mathcal{H}^{\text{init}}) = \bigcup_{r\in\text{appl}(\mathfrak{s})} \text{crit}(\mathcal{S}^{\text{in}}_{r(\mathfrak{s})}\langle MS_r\rangle) \subseteq \bigcup_{r\in\text{appl}(\mathfrak{s})} \hat{\mathcal{H}}^2_{r(\mathfrak{s})}$$

$$\mathrm{crit}(\mathcal{H}^1_{\mathfrak{p}_1}) = \mathrm{crit}(\mathcal{I}^{\mathrm{init}}_{\mathfrak{p}_1}) \cup \bigcup_{r \in \mathrm{appl}(\mathfrak{p}_1)} \mathrm{crit}(\mathcal{S}^{\mathrm{in}}_{r(\mathfrak{p}_1)}\langle MS_r \rangle) \subseteq \hat{\mathcal{H}}^{\mathrm{init}} \cup \bigcup_{r \in \mathrm{appl}(\mathfrak{p}_1)} \hat{\mathcal{H}}^2_{r(\mathfrak{p}_1)}$$

$$\mathrm{crit}(\mathcal{H}^2_{\mathfrak{p}_2}) = \mathrm{crit}(\mathcal{I}^{\mathrm{init}}_{\mathfrak{p}_2}) \cup \mathrm{crit}(\mathcal{S}^{\mathrm{in}}_{r_{\kappa(\mathfrak{p}_2)}(\mathfrak{p}_2)}\langle MD_{r_{\kappa(\mathfrak{p}_2)}} \rangle) \subseteq \hat{\mathcal{H}}^{\mathrm{init}} \cup \hat{\mathcal{H}}^1_{r_{\kappa(\mathfrak{p}_2)}(\mathfrak{p}_2)}$$

Since $\mathfrak{s} \in \mathcal{K}_1$, it follows that $r(\mathfrak{s}) \in \mathcal{K}_2$ and hence $\mathcal{H}^2_{r(\mathfrak{s})} \subseteq \mathcal{H}$ for all $r$ applicable to $\mathfrak{s}$. Because $\mathfrak{p}_1 \in \mathcal{K}_1$, it holds that $r(\mathfrak{p}_1) \in \mathcal{K}_2$ and thus $\mathcal{H}^2_{r(\mathfrak{p}_1)} \subseteq \mathcal{H}$ for all $r \in \mathrm{appl}(\mathfrak{p}_1)$. Since $\mathfrak{p}_2 \in \mathcal{K}_2$, it follows that $r_{\kappa(\mathfrak{p}_2)}(\mathfrak{p}_2) \in \mathcal{K}_1$ and thus $\mathcal{H}^1_{r_{\kappa(\mathfrak{p}_2)}(\mathfrak{p}_2)} \subseteq \mathcal{H}$. Consequently, all strategies mentioned in the above inclusion are contained in $\mathcal{H}$. ☐

*Proof of the First Main Lemma (Lemma 11).* It is easy to verify that the reduction can be performed in LOGSPACE. Lemma 11 then follows from Lemma 26 and Lemma 27. ☐

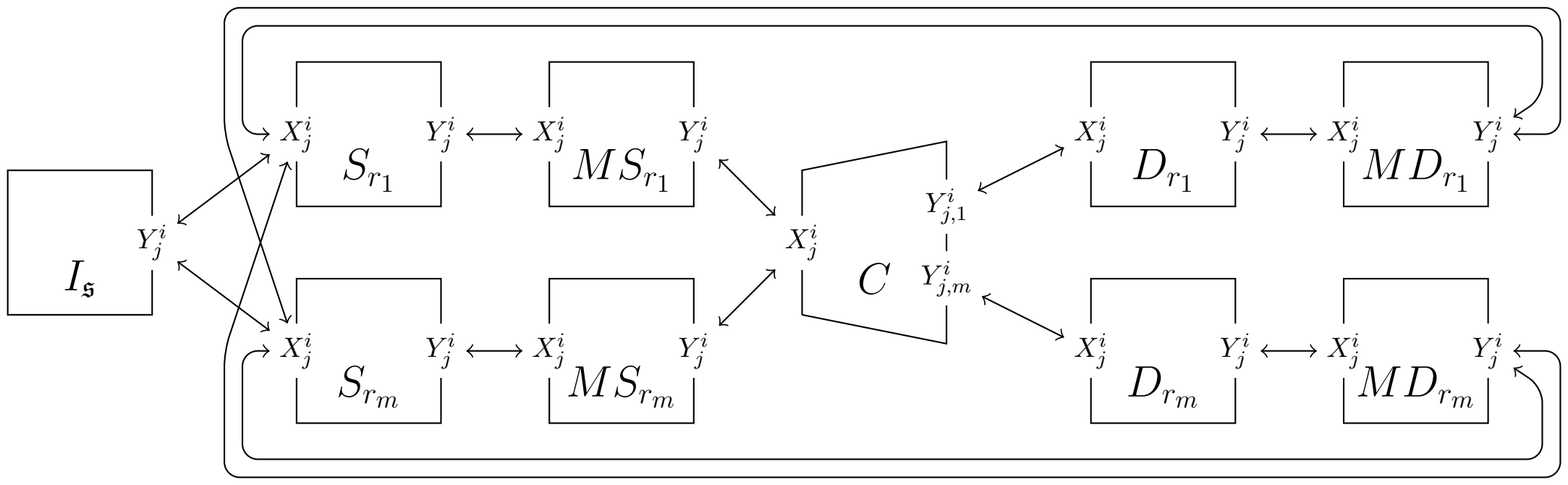

Figure 2: The 3-CNF formula $\Gamma'(G)$.

*Proof of the Second Main Lemma (Lemma 12).* The number of clauses used by all gadgets in $\Gamma(G)$ is bounded by $O(\|G\|^4)$. The most wasteful part is the set of $O(m)$ Choice Gadgets of size $O(knm^2)$ each. However, since we do not argue about regular refutations here, it suffices to use one Choice Gadget whose input variables are connected to the output variables of all $MS_r$ gadgets. The modified construction $\Gamma'(G)$ is illustrated in Figure 2. The proof of Lemma 26 and Lemma 27 goes through with that modification (except for regularity). With this clause set $\Gamma'(G)$ we get $\|\Gamma'(G)\| = O(\|G\|^3)$ and $|\mathrm{Var}(\Gamma'(G))| = O(\|G\|^2)$ as desired. ☐

## 6. The Length of the Narrowest Proof

Despite the hardness of even deciding the existence of a narrow proof, the minimum width heuristics performs in some cases better than the DPLL procedure. In order to compare the power of the two approaches one compares the length of a minimum width refutation with the length of a minimal treelike refutation. First, if $\Gamma$ has a treelike refutation of length $S$, then it has also a refutation of width $O(\log S)$ and hence a minimum width refutation of length $n^{O(\log S)}$ [4,6]. Thus, the length of the narrowest refutation is quasi-polynomial bounded in the length of a minimal treelike resolution refutation. Furthermore, Ben-Sasson, Impagliazzo and Widgerson [5] constructed a sequence of CNF formulas $\{\Gamma_n\}_{n=1}^{\infty}$ of size $\|\Gamma_n\| = O(n)$ such that:

- Every treelike resolution refutation of $\Gamma_n$ has length at least $2^{\Omega(\frac{n}{\log n})}$.
- There is a resolution refutation of $\Gamma_n$ of width $O(1)$.

This provides an example where the minimum width heuristics succeeds in polynomial time whereas every implementation of the DPLL procedure requires exponential time. Theorem 5 provides for every constant $k$ a contrary example: The minimum width heuristics produces refutations of length $\Omega(n^{k-1})$ and width $k$ while there exists a treelike refutation of constant length (depending on $k$) and width $k+1$. Therefore, the refutation of minimal width is by far longer than the treelike refutation of minimal length. We present a short proof sketch of Theorem 5 here, a detailed proof can be found in the appendix.

*Proof of Theorem 5 (sketch).* We use a similar construction as in the reduction above to prove a lower bound on the number of rounds in the width-$k$ game. The variable blocks $X^i_j$ for $i \in [k-1]$ and $j \in [n]$ were used to encode an $n$-ary counter with $k-1$ digits. Thus, every position $\mathfrak{p}$ on $X$ encodes a number from 0 to $n^{k-1}-1$. The start position is the position identified with 0 and $k-1$ Rule Gadgets were used to increment the counter. Spoiler wins if he reaches a position that corresponds to the number $(n-1)n^{k-2}$, but since he has to run through the gadgets and increment $\Omega(n^{k-1})$ times, this gives a lower bound on the number of rounds in the game. By Lemma 6 this is a lower bound on the depth and hence on the length of width-$k$ resolution refutations. □

# A. Strategies on the Switch

Let us start with some notation. Recall the clauses of the Switch (see Section 4.3), they contain the following variables.

$$X(M)^i_j, \qquad i \in [k], j \in [n]; \tag{30}$$
$$A0^i_j, A1^i_j, A^{i,1}_j, A^{i,2}_j, A^{i,3}_j, A^{i,4}_j, \qquad i \in [k], j \in [n]; \tag{31}$$
$$A^{i,c}_{j,l}, \qquad c \in [4], i \in [k], j \in [n], l \in [k]; \tag{32}$$
$$A^{i,c}_{j,\geq l}, \qquad c \in [4], i \in [k], j \in [n], 2 \leq l \leq k-1; \tag{33}$$
$$B_l, \qquad l \in [k]; \tag{34}$$
$$B_{\geq l}, \qquad 2 \leq l \leq k-1; \tag{35}$$
$$B, \tag{36}$$
$$Y(M)^i_j, \qquad i \in [k], j \in [n]. \tag{37}$$

Let $\mathsf{A}^i$ be the set of variables from lines (31) – (33) with upper index $i$ and $\mathsf{A} := \bigcup_{i \in [k]} \mathsf{A}^i$. By $\mathsf{B}$ we denote the set of variables from lines (34) – (36).

## A.1. Proof of Lemma 20

*Proof of Lemma 20.* Starting from $\{X^i_{\mathfrak{p}(i)} \mapsto 1 \mid i \in [k]\}$ Spoiler asks for $A0^1_{\mathfrak{p}(1)}$ and $A1^1_{\mathfrak{p}(1)}$. Because of (10), Duplicator has to answer 1 for $A0^1_{\mathfrak{p}(1)}$ or $A1^1_{\mathfrak{p}(1)}$. Assume that Spoiler reaches $A0^1_{\mathfrak{p}(1)} \mapsto 1$. Then he picks up the pebbles from $A1^1_{\mathfrak{p}(1)}$ and $X^1_{\mathfrak{p}(1)}$ and places them on $A^{1,1}_{\mathfrak{p}(1)}$ and $A^{1,2}_{\mathfrak{p}(1)}$. Owing to (11), Duplicator has to answer 1 for $A^{1,1}_{\mathfrak{p}(1)}$ or $A^{1,2}_{\mathfrak{p}(1)}$. If Duplicator has answered with $A1^1_{\mathfrak{p}(1)} \mapsto 1$ above, then Spoiler could reach $A^{1,3}_{\mathfrak{p}(1)} \mapsto 1$ or $A^{1,4}_{\mathfrak{p}(1)} \mapsto 1$ in an analog way. Thus, there is a $c \in [4]$ of Duplicator's choice such that Spoiler can reach $\{A^{1,c}_{\mathfrak{p}(1)} \mapsto 1\} \cup \{X^i_{\mathfrak{p}(i)} \mapsto 1 \mid 2 \leq i \leq k\}$. Spoiler now takes the remaining two pebbles and applies the same strategy to the other partitions. Therefore, he reaches the position $\{A^{i,b(i)}_{\mathfrak{p}(i)} \mapsto 1 \mid i \in [k]\}$ for some mapping $b : [k] \to [4]$. Like before, Spoiler uses one pebble to store information on one partition and two pebbles to walk through the partition. Using (13), (14) and (15) Spoiler can reach position $A^{1,b(1)}_{\mathfrak{p}(1),l_1} \mapsto 1$, for some $l_1 \in [k]$ of Duplicator's choice, without grabbing the pebbles from the other partitions. Using the same technique Spoiler can reach $A^{2,b(2)}_{\mathfrak{p}(2),l_2} \mapsto 1$ in partition 2. If Duplicator chooses $l_2 = l_1$, then she looses immediately because of clause (16). Thus, she has to choose some $l_2 \in [k] \setminus \{l_1\}$. Following that strategy, Spoiler reaches $A^{i,b(i)}_{\mathfrak{p}(i),l_i} \mapsto 1$ for some $l_i \in [k]\setminus\{l_1,\ldots,l_{i-1}\}$ and eventually $\{A^{i,b(i)}_{\mathfrak{p}(i),\sigma(i)} \mapsto 1 \mid i \in [k]\}$ for some permutation $\sigma\colon [k] \to [k]$.

At this point Spoiler asks for $B$. Assume first that Duplicator answers with 1. Because of clause (21) Spoiler can force Duplicator to answer 1 when he asks for $Y^1_{\mathfrak{p}(1)}$ with the remaining pebble. Then Spoiler picks up the pebble from $A^{1,b(1)}_{\mathfrak{p}(1),\sigma(1)}$ and puts it on $Y^2_{\mathfrak{p}(2)}$. Once more, clause (21) forces Duplicator to answer with 1. Playing that strategy also on the other partitions, Spoiler can reach $\{Y^i_{\mathfrak{p}(i)} \mapsto 1 \mid i \in [k]\}$ and is done.

So assume that Duplicator answers 0 when he is asked for $B$. The current position is $\{A^{i,b(i)}_{\mathfrak{p}(i),\sigma(i)} \mapsto 1 \mid i \in [k]\} \cup \{B \mapsto 0\}$. Using the clauses (17) and the $(k+2)$-th pebble, Spoiler can reach $\{B_l \mapsto 1 \mid l \in [k]\} \cup \{B \mapsto 0\}$. In the next step Spoiler asks for $B_{\geq k-1}$ and Duplicator has to answer with 1 owing to clause (20). Then Spoiler picks up the pebble from $B_k$ and asks for $B_{\geq k-2}$. Duplicator has to answer with 1 according to clause (19). Following that strategy Spoiler can reach positions $\{B_l \mapsto 1 \mid 1 \leq l < i\} \cup \{B_{\geq i} \mapsto 1, B \mapsto 0\}$ for $i = (k-1)\ldots 2$. Since $\{B_1 \mapsto 1, B_{\geq 2} \mapsto 1, B \mapsto 0\}$ falsifies clause (18), Spoiler wins the game. ☐

## A.2. Proof of Lemma 21

*Proof of Lemma 21.* Let $T \subseteq [k]$ be a nonempty set and $\mathfrak{p} : [k] \to [n]$ a position. We define a total assignment $S^{\mathrm{imp}}_{\mathfrak{p},T}$ that falsifies no clause from the Switch. Since $T$ is nonempty we can fix some $t^* \in T$, say the minimal one.

$$X^i_j \mapsto \begin{cases} 1, \text{ if } i \notin T \text{ and } j = \mathfrak{p}(i) \\ 0, \text{ else} \end{cases} \tag{38}$$

$$A0^i_j \mapsto \begin{cases} 1, \text{ if } i \notin T \text{ and } j = \mathfrak{p}(i) \\ 0, \text{ else} \end{cases} \tag{39}$$

$$A1^i_j \mapsto 0 \tag{40}$$

$$A^{i,c}_j \mapsto \begin{cases} 1, \text{ if } i \notin T,\ j = \mathfrak{p}(i) \text{ and } c = 1 \\ 0, \text{ else} \end{cases} \tag{41}$$

$$A^{i,c}_{j,l} \mapsto \begin{cases} 1, \text{ if } i \notin T,\ j = \mathfrak{p}(i),\ c = 1 \text{ and } l = i \\ 0, \text{ else} \end{cases} \tag{42}$$

$$A^{i,c}_{j,\geq l} \mapsto \begin{cases} 1, \text{ if } i \notin T,\ j = \mathfrak{p}(i),\ c = 1 \text{ and } l \leq i \\ 0, \text{ else} \end{cases} \tag{43}$$

$$B_l \mapsto \begin{cases} 1, \text{ if } l \notin T \\ 0, \text{ else} \end{cases} \tag{44}$$

$$B_{\geq l} \mapsto \begin{cases} 1, \text{ if } l > t^* \\ 0, \text{ else} \end{cases} \tag{45}$$

$$B \mapsto 0 \tag{46}$$

$$Y^i_j \mapsto 0 \tag{47}$$

For all positions $\mathfrak{p} : [k] \to [n]$, mappings $b : [k] \to [4]$ and a permutations $\sigma : [k] \to [k]$ the satisfying assignment $S^{\mathrm{out}}_{\mathfrak{p},b,\sigma}$ is defined as follows.

$$X^i_j \mapsto 0 \tag{48}$$

$$A0^i_j \mapsto \begin{cases} 1, \text{ if } j = \mathfrak{p}(i) \text{ and } b(i) \in \{1,2\} \\ 0, \text{ else} \end{cases} \tag{49}$$

$$A1^i_j \mapsto \begin{cases} 1, \text{ if } j = \mathfrak{p}(i) \text{ and } b(i) \in \{3,4\} \\ 0, \text{ else} \end{cases} \tag{50}$$

$$A^{i,c}_j \mapsto \begin{cases} 1, \text{ if } j = \mathfrak{p}(i) \text{ and } b(i) = c \\ 0, \text{ else} \end{cases} \tag{51}$$

$$A^{i,c}_{j,l} \mapsto \begin{cases} 1, \text{ if } j = \mathfrak{p}(i),\ b(i) = c \text{ and } l = \sigma(i) \\ 0, \text{ else} \end{cases} \tag{52}$$

$$A^{i,c}_{j,\geq l} \mapsto \begin{cases} 1, \text{ if } j = \mathfrak{p}(i),\ b(i) = c \text{ and } l \leq \sigma(i) \\ 0, \text{ else} \end{cases} \tag{53}$$

$$B_l \mapsto 1 \tag{54}$$

$$B_{\geq l} \mapsto 1 \tag{55}$$

$$B \mapsto 1 \tag{56}$$

$$Y^i_j \mapsto \begin{cases} 1, \text{ if } j = \mathfrak{p}(i) \\ 0, \text{ else.} \end{cases} \tag{57}$$

Let $t \in [k]$, $\mathfrak{p} : [k] \to [n]$, $b : [k] \to [4]$ and $\sigma : [k] \to [k]$ be a permutation on $[k]$. We define the following partial assignment $S^t_{\mathfrak{p},b,\sigma}$ (that will later be used for the input strategies).

$$X^i_j \mapsto \begin{cases} 1, \text{ if } j = \mathfrak{p}(i) \\ 0, \text{ else} \end{cases} \tag{58}$$

$$A0^i_j \mapsto \begin{cases} 1, \text{ if } j = \mathfrak{p}(i) \text{ and } b(i) \in \{1,2\} \\ 0, \text{ else} \end{cases} \tag{59}$$

$$A1^i_j \mapsto \begin{cases} 1, \text{ if } j = \mathfrak{p}(i) \text{ and } b(i) \in \{3,4\} \\ 0, \text{ else} \end{cases} \tag{60}$$

$$A^{i,c}_j \mapsto \begin{cases} \text{undefined, if } i = t,\ j = \mathfrak{p}(i) \text{ and } c = b(i) \\ 1, \text{ if } i \neq t,\ j = \mathfrak{p}(i) \text{ and } c = b(i) \\ 0, \text{ else} \end{cases} \tag{61}$$

$$A^{i,c}_{j,l} \mapsto \begin{cases} \text{undefined, if } i = t,\ j = \mathfrak{p}(i),\ c = b(i) \text{ and } l = \sigma(i) \\ 1, \text{ if } i \neq t,\ j = \mathfrak{p}(i),\ c = b(i) \text{ and } l = \sigma(i) \\ 0, \text{ else} \end{cases} \tag{62}$$

$$A^{i,c}_{j,\geq l} \mapsto \begin{cases} \text{undefined, if } i = t,\ j = \mathfrak{p}(i),\ c = b(i) \text{ and } l \leq \sigma(i) \\ 1, \text{ if } i \neq t,\ j = \mathfrak{p}(i),\ c = b(i) \text{ and } l \leq \sigma(i) \\ 0, \text{ else} \end{cases} \tag{63}$$

$$B_l \mapsto \begin{cases} 0, \text{ if } l = \sigma(t) \\ 1, \text{ else} \end{cases} \tag{64}$$

$$B_{\geq l} \mapsto \begin{cases} 0, \text{ if } l \leq \sigma(t) \\ 1, \text{ else} \end{cases} \tag{65}$$

$$B \mapsto 0 \tag{66}$$

$$Y^i_j \mapsto 0 \tag{67}$$

For the impasse and output strategies we simply take the set of all partial assignments contained in the total assignments as winning strategies. For the input strategies we have to be more careful and set $\mathcal{S}^t_{\mathfrak{p},b,\sigma} = \{p \mid p \subset S^t_{\mathfrak{p},b,\sigma}, |\operatorname{Dom}(p)| \leq k+2, |\operatorname{Dom}(p) \cap \mathsf{A}^t| \leq 2\}$.

$$\mathcal{S}^{\text{out}}_{\mathfrak{p}} := \{p \mid p \subseteq S^{\text{out}}_{\mathfrak{p},b,\sigma}, b\colon [k] \to [4], \sigma\colon [k] \to [k]\} \tag{68}$$

$$\mathcal{S}^{\text{imp}}_{\mathfrak{p},T} := \{p \mid p \subseteq S^{\text{imp}}_{\mathfrak{p},T}\} \tag{69}$$

$$\mathcal{S}^{\text{in}}_{\mathfrak{p}} := \bigcup_{t;\, b;\, \sigma} \mathcal{S}^{t}_{\mathfrak{p},b,\sigma} \tag{70}$$

First we fix some nonempty $T \subseteq [k]$ and position $\mathfrak{p}\colon [k] \to [n]$. It is clear by definition that all strategies have the desired boundary and satisfy the closure property. Furthermore, the output strategies $\mathcal{S}^{\text{out}}_{\mathfrak{p}}$ and the impasse strategies $\mathcal{S}^{\text{imp}}_{\mathfrak{p},T}$ are winning strategies since they are drawn from total assignments that falsify no clause.

The most difficult case is (iii). We define $\text{crit}(\mathcal{S}^{\text{in}}_{\mathfrak{p}})$ to be the set of all positions $p \in \mathcal{S}^{\text{in}}_{\mathfrak{p}}$ with domain size $k+1$ such that there exists a $t \in [k]$ with $|\operatorname{Dom}(p) \cap \mathsf{A}^t| = 2$ and $|\operatorname{Dom}(p) \cap \mathsf{A}^i| = 1$ for all $i \in [k] \setminus \{t\}$. For every $p \in \text{crit}(\mathcal{S}^{\text{in}}_{\mathfrak{p}})$ it holds that $\operatorname{Dom}(p) \subseteq \mathsf{A}$ and $p \subseteq S^{t}_{\mathfrak{p},b,\sigma}$ for some $b\colon [k] \to [4]$ and $\sigma\colon [k] \to [k]$. It follows that $p \subseteq S^{\text{out}}_{\mathfrak{p},b,\sigma}$ and hence $\text{crit}(\mathcal{S}^{\text{in}}_{\mathfrak{p}}) \subseteq \mathcal{S}^{\text{out}}_{\mathfrak{p}}$. It remains to show that all $p \in \mathcal{S}^{\text{in}}_{\mathfrak{p}} \setminus \text{crit}(\mathcal{S}^{\text{in}}_{\mathfrak{p}})$ satisfy the extension property. We can fix $t, b, \sigma$ such that $p \subset S^{t}_{\mathfrak{p},b,\sigma}$, $|\operatorname{Dom}(p)| \le k+1$, and $|\operatorname{Dom}(p) \cap \mathsf{A}^t| \le 2$. If Spoiler asks for a variable $Z \notin \mathsf{A}^t$, then Duplicator answers with $z \in \{0,1\}$ such that $p \cup \{Z \mapsto z\} \in \mathcal{S}^{t}_{\mathfrak{p},b,\sigma}$. So we can assume that $Z \in \mathsf{A}^t$.

**Case 1:** $|\operatorname{Dom}(p) \cap \mathsf{A}^t| \le 1$.

If $Z \in \operatorname{Dom}(S^{t}_{\mathfrak{p},b,\sigma})$, then Duplicator can answer with $z \in \{0,1\}$ such that $p \cup \{Z \mapsto z\} \in \mathcal{S}^{t}_{\mathfrak{p},b,\sigma}$ because $|\operatorname{Dom}(p \cup \{Z \mapsto z\}) \cap \mathsf{A}^t| \le 2$. Thus, we can assume that $Z \in \mathsf{A}^t \setminus \operatorname{Dom}(S^{t}_{\mathfrak{p},b,\sigma})$.

**Case 1.1:** $\{A0^{t}_{\mathfrak{p}(t)}, A1^{t}_{\mathfrak{p}(t)}\} \cap \operatorname{Dom}(p) = \emptyset$.

We can fix a $c \in [4]$ such that the variable in $\operatorname{Dom}(p) \cap \mathsf{A}^t$ is one of the variables $A^{t,c}_{\mathfrak{p}(t)}$, $A^{t,c}_{\mathfrak{p}(t),\ge l}$ or $A^{t,c}_{\mathfrak{p}(t),l}$. If $\operatorname{Dom}(p) \cap \mathsf{A}^t = \emptyset$ we set $c = 1$. Now we can flip $b$ on partition $t$:

$$\hat{b}(t) := \begin{cases} 1, \text{ if } b(t) \in \{3,4\} \text{ and } c \ne 1, \\ 2, \text{ if } b(t) \in \{3,4\} \text{ and } c = 1, \\ 3, \text{ if } b(t) \in \{1,2\} \text{ and } c \ne 3, \\ 4, \text{ if } b(t) \in \{1,2\} \text{ and } c = 3. \end{cases}$$

We set $\hat{b}(i) := b(i)$ for all other $i \in [k] \setminus \{t\}$. It follows $p \cup \{Z \mapsto 0\} \in \mathcal{S}^{t}_{\mathfrak{p},\hat{b},\sigma}$ for all $Z \in \mathsf{A}^t \setminus \operatorname{Dom}(S^{t}_{\mathfrak{p},b,\sigma})$.

**Case 1.2:** $\operatorname{Dom}(p) \cap \mathsf{A}^t \subset \{A0^{i}_{\mathfrak{p}(i)}, A1^{i}_{\mathfrak{p}(i)}\}$.

Once more we can flip $b$ to $\hat{b}$ ensuring that $p \cup \{Z \mapsto 0\} \in \mathcal{S}^{t}_{\mathfrak{p},\hat{b},\sigma}$. We let $\hat{b}(i) := b(i)$ for all $i \in [k] \setminus \{t\}$ and

$$\hat{b}(t) := \begin{cases} 1, \text{ if } b(i) = 2, \\ 2, \text{ if } b(i) = 1, \\ 3, \text{ if } b(i) = 4, \\ 4, \text{ if } b(i) = 3. \end{cases}$$

**Case 2:** $|\operatorname{Dom}(p) \cap \mathsf{A}^t| = 2$.

**Case 2.1:** $\operatorname{Dom}(p) \cap \mathsf{B} \ne \emptyset$.

Since $|\operatorname{Dom}(p)| \le k+1$, there is some partition $j$ such that $\operatorname{Dom}(p) \cap \mathsf{A}^j = \emptyset$. We define new parameters $\hat{t}$, $\hat{\sigma}$, $\hat{b}$ such that $p$ is contained in $\mathcal{S}^{\hat{t}}_{\mathfrak{p},\hat{b},\hat{\sigma}}$. Let $\hat{t} := j$, $\hat{\sigma}(i) := \sigma(i)$ for $i \in [k] \setminus \{j,t\}$, $\hat{\sigma}(t) := \sigma(j)$ and $\hat{\sigma}(j) := \sigma(t)$. Furthermore, we define $\hat{b}(i) := b(i)$ for $i \in [k] \setminus \{t\}$ and will define $\hat{b}(t)$ in the sequel. Note that all variables but those from partition $t$ are mapped to the same value in $S^t_{\mathfrak{p},b,\sigma}$ as they were mapped to in $S^{\hat{t}}_{\mathfrak{p},\hat{b},\hat{\sigma}}$, independent of the choice of $\hat{b}(t)$. Especially, all variables in $\mathsf{B}$ stay the same since $\sigma(t) = \hat{\sigma}(\hat{t})$. Furthermore, $|\,\mathrm{Dom}(p) \cap \mathsf{A}^{\hat{t}}\,| = 0 \leq 2$. Since $\mathcal{S}^{\hat{t}}_{\mathfrak{p},\hat{b},\hat{\sigma}}$ is defined on all variables in $\mathsf{A}^t$, Duplicator can always provide an answer $z$ for $Z$ such that $p \cup \{Z \mapsto z\} \in \mathcal{S}^{\hat{t}}_{\mathfrak{p},\hat{b},\hat{\sigma}}$. It remains to show that $p$ restricted to $\mathsf{A}^t$ is a subset of $S^{\hat{t}}_{\mathfrak{p},\hat{b},\hat{\sigma}}$. We establish this fact by flipping $b(t)$ to $\hat{b}(t)$ as follows.
**Case 2.1.1:** $\{A0^t_{\mathfrak{p}(t)}, A1^t_{\mathfrak{p}(t)}\} \cap \mathrm{Dom}(p) = \emptyset$.

In this case $\hat{b}(t)$ is defined to be the smallest $c \in [4]$ such that there is no pebble on a variable of the form $A^{t,c}_{\mathfrak{p}(t)}$, $A^{t,c}_{\mathfrak{p}(t),l}$ or $A^{t,c}_{\mathfrak{p}(t),\geq l}$. Such a $c$ exists since there are exactly two pebbles in $\mathrm{Dom}(p) \cap \mathsf{A}^t$.
**Case 2.1.2:** $\{A0^t_{\mathfrak{p}(t)}, A1^t_{\mathfrak{p}(t)}\} \cap \mathrm{Dom}(p) \neq \emptyset$.

One of the two pebbles from $\mathrm{Dom}(p) \cap \mathsf{A}^t$ is on $A0^t_{\mathfrak{p}(t)}$ or $A1^t_{\mathfrak{p}(t)}$. If the other pebble is not on some variable $A^{t,b(t)}_{\mathfrak{p}(t)}$, $A^{t,b(t)}_{\mathfrak{p}(t),l}$ or $A^{t,b(t)}_{\mathfrak{p}(t),\geq l}$ (for some $l \in [k]$) we let $\hat{b}(t) := b(t)$. Otherwise,

$$\hat{b}(t) := \begin{cases} 1, \text{ if } b(t) = 2, \\ 2, \text{ if } b(t) = 1, \\ 3, \text{ if } b(t) = 4, \\ 4, \text{ if } b(t) = 3. \end{cases}$$

**Case 2.2:** $\mathrm{Dom}(p) \cap \mathsf{B} = \emptyset$.
**Case 2.2.1:** There exists a $j \in [k]$ such that $\mathrm{Dom}(p) \cap \mathsf{A}^j = \emptyset$.

There is no pebble on $\mathsf{B}$ and no pebble on $\mathsf{A}^j$, therefore $p$ is also contained in $\mathcal{S}^j_{\mathfrak{p},b,\sigma}$. Since $\mathsf{A}^t \subseteq \mathrm{Dom}(S^j_{\mathfrak{p},b,\sigma})$, Duplicator can provide an answer $z$ for every requested $Z \in \mathsf{A}^t$ such that $p \cup \{Z \mapsto z\} \in \mathcal{S}^j_{\mathfrak{p},b,\sigma}$.
**Case 2.2.2:** For all $i \in [k]$: $|\,\mathrm{Dom}(p) \cap \mathsf{A}^i\,| \geq 1$.

In this case $p \in \mathrm{crit}(\mathcal{S}^{\mathrm{in}}_{\mathfrak{p}})$ and there is nothing to show. ☐

# B. Proof of the Lower Bound on the Length of Bounded Width Refutations

To prove Theorem 5 we use a similar construction as in the reduction above. Since the gadgets from the reduction are designed for the width-$(k+1)$ game let us restate Theorem 5 in terms of width-$(k+1)$ resolution (just for convenience).

**Theorem 5'.** *For every fixed integer $k \geq 2$ there is a family of unsatisfiable 3-CNF formulas $\{\Gamma^k_n\}^\infty_{n=1}$ with $O(n)$ variables, $O(n^2)$ clauses and minimal refutation width $k+1$ such that the following holds:*

- *Every width-$(k+1)$ resolution refutation of $\Gamma^k_n$ has length at least $\Omega(n^k)$.*
- *There is a width-$(k+2)$ treelike resolution refutation of $\Gamma^k_n$ of length $O(1)$.*

*Proof.* The 3-CNF $\Gamma^k_n$ uses the same Switches $M$ and the same Initialization Gadget $I_{\mathfrak{s}}$ as in the reduction from the $k$-pebble KAI game with $|U| = n$ and Increment Gadgets (defined below) similar to the Rule Gadgets for Spoiler. The idea is to implement an $n$-ary counter with $k$ digits. We provide a bijection $\alpha$ between the (not necessarily injective) mappings $\mathfrak{p} : [k] \to [n]$ and the numbers $0, \dots, n^k - 1$ by setting $\alpha(\mathfrak{p}) = \sum_{i=0}^{k-1} (\mathfrak{p}(i+1) - 1)n^i$. Thus, $\mathfrak{p}(i) - 1$ is the $i$-th digit of the $n$-ary counter. We use the Initialization Gadget with start position $\mathfrak{s} = \alpha^{-1}(0)$ and connect the output to the variables $\{\mathsf{X}^i_j \mid i \in [k], j \in [n]\}$. We introduce the clause $\{\neg \mathsf{X}^k_n\}$ to ensure that

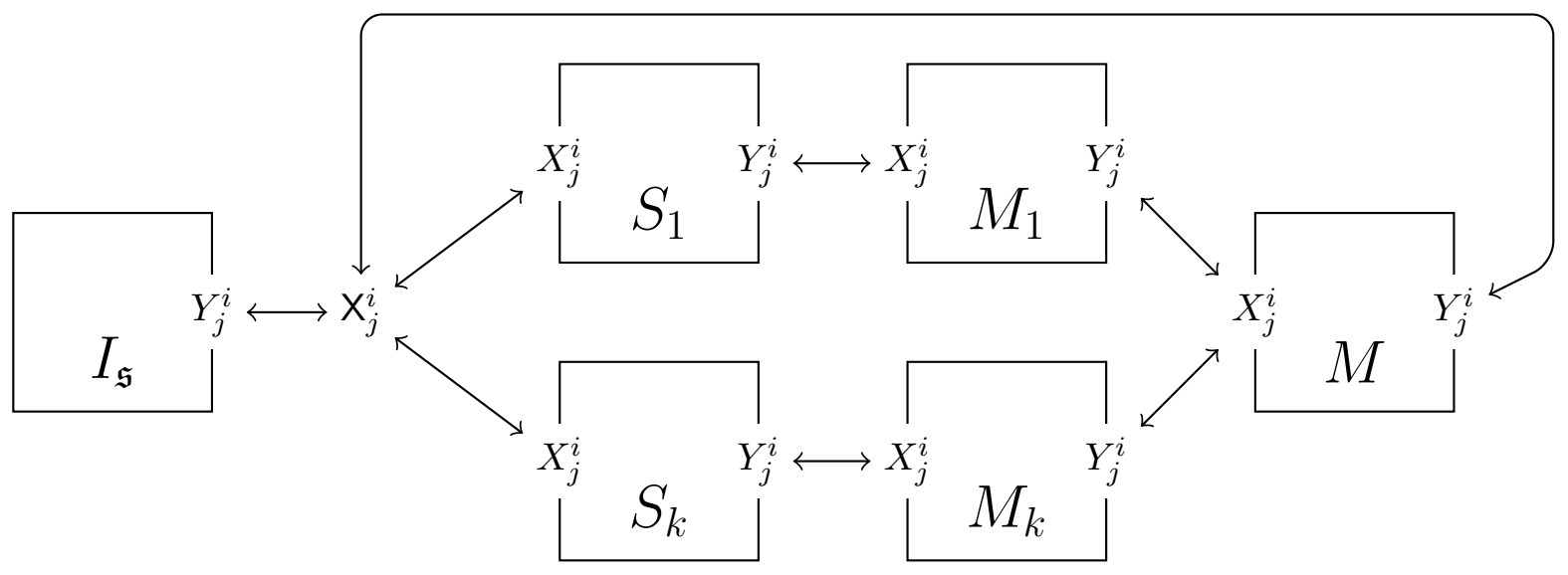

Figure 3: The 3-CNF formula $\Gamma^k_n$.

Spoiler wins, if he has stored the position $\{\mathsf{X}^i_{\mathfrak{p}(i)} \mapsto 1 \mid i \in [k]\}$ for some mapping $\mathfrak{p}$ satisfying $\alpha(\mathfrak{p}) \geq (n-1)n^{k-1}$. Furthermore, there is a clause $\{\neg\mathsf{X}^1_1, \neg\mathsf{X}^1_2\}$ that we will need later. We have to define Increment Gadgets in such a way that Spoiler can reach $\{\mathsf{X}^i_{\mathfrak{p}'(i)} \mapsto 1 \mid i \in [k]\}$ from $\{\mathsf{X}^i_{\mathfrak{p}(i)} \mapsto 1 \mid i \in [k]\}$ if $\alpha(\mathfrak{p}') = \alpha(\mathfrak{p}) + 1$. We introduce $k$ Increment Gadgets $S_1, \dots, S_k$ to perform this and connect the inputs of all Increment Gadgets to $\{\mathsf{X}^i_j \mid i \in [k], j \in [n]\}$. The output of every Increment Gadget $S_l$ is connected to the input of an Switch $M_l$. The outputs of all Switches $M_1, \dots, M_k$ are connected to the input of one additional Switch $M$ and the output of $M$ is connected to $\{\mathsf{X}^i_j \mid i \in [k], j \in [n]\}$. Starting from the $\mathsf{X}$-vertices Spoiler can move through some Increment Gadget $S_l$ to increment the counter and then he moves through the Switches $M_l$ and $M$ to reach the new incremented position at the $\mathsf{X}$-vertices. Now we define the Increment Gadget $S_l$. Just like the Rule Gadgets, the Increment Gadget consists of input vertices $X(S_l)^i_j$ and output vertices $Y(S_l)^i_j$ for all $i \in [k]$ and $j \in [n]$. There are the following clauses:

$$X(S_l)^i_n \to Y(S_l)^i_1 \qquad \text{for all } i < l \tag{71}$$

$$X(S_l)^l_j \to Y(S_l)^l_{j+1} \qquad \text{for all } 1 \leq j < n \tag{72}$$

$$X(S_l)^i_j \to Y(S_l)^i_j \qquad \text{for all } i > l \text{ and } j \in [n] \tag{73}$$

It follows that Spoiler can reach $\{Y(S_l)^i_{\mathfrak{p}'(i)} \mapsto 1 \mid i \in [k]\}$ from $\{X(S_l)^i_{\mathfrak{p}(i)} \mapsto 1 \mid i \in [k]\}$ if $\alpha(\mathfrak{p}) = a_{k-1}n^{k-1} + \dots + a_{l+1}n^{l+1} + a_l n^l + (n-1)n^{l-1} + \dots + (n-1)n^0$ with $a_l < n-1$ and $\alpha(\mathfrak{p}') = a_{k-1}n^{k-1} + \dots + a_{l+1}n^{l+1} + (a_l+1)n^l$. Hence, $\alpha(\mathfrak{p}') = \alpha(\mathfrak{p}) + 1$. Thus, Spoiler can always choose an Increment Gadget to increase the current position by one. On the other hand, Duplicator has a strategy $\mathcal{R}_\mathfrak{p}$ such that Spoiler can not reach a valid position at the output, if he chooses a false Increment Gadget. Let $T_l(\mathfrak{p})$ be the set $\{i \in [l] \mid (i < l \text{ and } \mathfrak{p}(i) \neq n) \text{ or } (i = l \text{ and } \mathfrak{p}(l) = n)\}$ of partitions witnessing that the Increment Gadget $S_l$ cannot be applied. For every mapping $\mathfrak{p}$ with $\mathfrak{p}(k) < n$ there is one unique $l_\mathfrak{p} \in [k]$ such that $T_{l_\mathfrak{p}}(\mathfrak{p}) = \emptyset$. Furthermore we define $\mathrm{inc}_l(\mathfrak{p}) : [k] \to [n]$ as follows

$$\mathrm{inc}_l(\mathfrak{p}) = \begin{cases} 1, \text{ if } i < l, \\ n, \text{ if } i = l, \\ \mathfrak{p}(i), \text{ if } i > l. \end{cases}$$

Thus, we get $\alpha(\mathrm{inc}_{l_\mathfrak{p}}(\mathfrak{p})) = \alpha(\mathfrak{p}) + 1$. $\mathcal{R}^l_\mathfrak{p}$ is the winning strategy for Duplicator on $S_l$ with boundary

$$\beta_\mathfrak{p}(X(S_l)^i_j) = \begin{cases} 1, \text{ if } j = \mathfrak{p}(i) \\ 0, \text{ else,} \end{cases} \qquad \beta_\mathfrak{p}(Y(S_l)^i_j) = \begin{cases} 1, \text{ if } j \notin T_l(\mathfrak{p}) \text{ and } j = \mathrm{inc}_l(\mathfrak{p})(i) \\ 0, \text{ else.} \end{cases}$$

Moreover, $\mathcal{R}_0^l$ is the strategy on $S_l$ that maps all variables to 0. For all $b = 1 \dots (n-1)n^{k-1} - 1$ we define the critical strategies $\mathcal{H}_b^1$, $\mathcal{H}_b^2$ and $\mathcal{H}^{\text{init}}$:

$$\mathcal{H}_b^1 := \mathcal{I}_{\mathfrak{p}}^{\text{init}} \uplus \biguplus_{l\in[k]} \mathcal{R}_{\mathfrak{p}}^l \uplus \mathcal{S}_{\alpha^{-1}(b+1)}^{\text{in}}\langle M_{l_\mathfrak{p}}\rangle \uplus \biguplus_{l\in[k]\setminus\{l_\mathfrak{p}\}} \mathcal{S}_{\text{inc}_l(\mathfrak{p}),T_l(\mathfrak{p})}^{\text{imp}}\langle M_l\rangle \uplus \mathcal{S}_{\mathfrak{p}}^{\text{out}}\langle M\rangle \quad (\mathfrak{p} := \alpha^{-1}(b))$$

$$\mathcal{H}_b^2 := \mathcal{I}_0^{\text{init}} \uplus \biguplus_{l\in[k]} \mathcal{R}_0^l \uplus \biguplus_{l\in[k]} \mathcal{S}_{\alpha^{-1}(b+1)}^{\text{out}}\langle M_l\rangle \uplus \mathcal{S}_{\alpha^{-1}(b+1)}^{\text{in}}\langle M\rangle$$

$$\mathcal{H}^{\text{init}} := \mathcal{I}^{\text{init}} \uplus \biguplus_{l\in[k]} \mathcal{R}_{\mathfrak{s}}^l \uplus \mathcal{S}_{\alpha^{-1}(1)}^{\text{in}}\langle M_{l_\mathfrak{s}}\rangle \uplus \biguplus_{l\in[k]\setminus\{l_\mathfrak{s}\}} \mathcal{S}_{\text{inc}_l(\mathfrak{s}),T_l(\mathfrak{s})}^{\text{imp}}\langle M_l\rangle \uplus \mathcal{S}_{\mathfrak{s}}^{\text{out}}\langle M\rangle \quad (\mathfrak{s} := \alpha^{-1}(0))$$

Now it holds that

$$\begin{aligned}\text{crit}(\mathcal{H}^{\text{init}}) &\subseteq \mathcal{H}_1^2 \setminus \text{crit}(\mathcal{H}_1^2)\\ \text{crit}(\mathcal{H}_b^1) &\subseteq \mathcal{H}_b^2 \setminus \text{crit}(\mathcal{H}_b^2) \cup \mathcal{H}^{\text{init}} \setminus \text{crit}(\mathcal{H}^{\text{init}})\\ \text{crit}(\mathcal{H}_b^2) &\subseteq \mathcal{H}_{b+1}^1 \setminus \text{crit}(\mathcal{H}_{b+1}^1) \cup \mathcal{H}^{\text{init}} \setminus \text{crit}(\mathcal{H}^{\text{init}})\end{aligned}$$

Duplicator can start with playing the critical strategy $\mathcal{H}^{\text{init}}$ and then switch to $\mathcal{H}_1^2$ as soon as Spoiler reaches some critical position. Then she plays according to $\mathcal{H}_1^2$ and switches to $\mathcal{H}_2^1$ if Spoiler reaches some critical position there. Following that strategy Duplicator stays always in the current critical strategy unless Spoiler reaches some critical position. Therefore, the only chance for Spoiler to win is to force Duplicator to flip the strategies $\mathcal{H}^{\text{init}}, \mathcal{H}_1^2, \mathcal{H}_2^1$, $\mathcal{H}_2^2$, $\mathcal{H}_3^1, \mathcal{H}_3^2$, $\dots$, $\mathcal{H}_b^2$ for $b = (n-1)n^{k-1} - 1$. Thus, he has to reach $2(n-1)n^{k-1} - 1$ critical positions and needs $\Omega(n^k)$ steps to win. It follows by Lemma 6 that every width-$k$ resolution refutation has depth at least $\Omega(n^k)$ and hence size at least $\Omega(n^k)$. Summing up the size of the gadgets yields that the formula $\Gamma_n^k$ contains $O(k^3 n)$ variables and $O(k^4 n^2)$ clauses.

Note that there is no resolution refutation of width $k$ since $\mathcal{H}^{\text{init}}$ is a winning strategy in the width-$k$ game (because every critical strategy in the width-$(k+1)$ game is a winning strategy in the width-$k$ game). On the other hand, in the width-$(k+2)$ game Spoiler can reach $\mathsf{X}_1^1 \mapsto 1$, store it and use the remaining pebbles to reach $\{\mathsf{X}_2^1 \mapsto 1\} \cup \{\mathsf{X}_1^i \mapsto 1 \mid 2 \le i \le k\}$ as in the width-$(k+1)$ game by incrementing once. Since he has stored $\mathsf{X}_1^1 \mapsto 1$ and $\mathsf{X}_2^1 \mapsto 1$ this falsifies clause $\{\neg\mathsf{X}_1^1, \neg\mathsf{X}_2^1\}$ introduced above. By looking into Spoilers strategies on the gadgets one can verify that this can be done within poly$(k)$ steps. Hence, there is a resolution refutation of depth poly$(k)$ by Lemma 6. It follows that there is a treelike resolution refutation of size $2^{\text{poly}(k)}$ which is constant if $k$ is fixed. ☐

## References

[1] A. Adachi, S. Iwata, and T. Kasai, "Some combinatorial game problems require $\Omega(n^k)$ time," *J. ACM*, vol. 31, Mar. 1984.

[2] A. Atserias and V. Dalmau, "A combinatorial characterization of resolution width," *J. Comput. Syst. Sci.*, vol. 74, no. 3, pp. 323–334, May 2008.

[3] A. Atserias, J. K. Fichte, and M. Thurley, "Clause-learning algorithms with many restarts and bounded-width resolution," *J. Artif. Intell. Res. (JAIR)*, vol. 40, pp. 353–373, 2011.

[4] P. Beame and T. Pitassi, "Simplified and improved resolution lower bounds," in *Proc. FOCS'06*, 1996, pp. 274–282.

[5] E. Ben-Sasson, R. Impagliazzo, and A. Wigderson, "Near optimal separation of tree-like and general resolution," *Combinatorica*, vol. 24, no. 4, Sep. 2004.

[6] E. Ben-Sasson and A. Wigderson, "Short proofs are narrow - resolution made simple," *J. ACM*, vol. 48, no. 2, pp. 149–169, 2001.

[7] C. Berkholz, "Lower bounds for existential pebble games and k-consistency tests," in *Proc. LICS'12*, 2012.

[8] A. K. Chandra, D. C. Kozen, and L. J. Stockmeyer, "Alternation," *J. ACM*, vol. 28, no. 1, pp. 114–133, Jan. 1981.

[9] Z. Galil, "On resolution with clauses of bounded size," *SIAM Journal on Computing*, vol. 6, no. 3, pp. 444–459, 1977.

[10] M. Grohe, "Equivalence in finite-variable logics is complete for polynomial time," in *Proc. FOCS'96*, 1996, pp. 264–273.

[11] A. Haken, "The intractability of resolution," *Theoretical Computer Science*, vol. 39, no. 0, pp. 297 – 308, 1985.

[12] A. Hertel and A. Urquhart, "The resolution width problem is EXPTIME-Complete," Tech. Rep. 133, 2006. [Online]. Available: http://eccc.hpi-web.de/report/2006/133/

[13] ——, "Comments on ECCC report TR06-133: the resolution width problem is EXPTIME-Complete," Tech. Rep. 003, 2009. [Online]. Available: http://eccc.hpi-web.de/report/2009/003/

[14] A. Hertel, "Applications of games to propositional proof complexity," Ph.D. dissertation, University of Toronto, 2008.

[15] T. Kasai, A. Adachi, and S. Iwata, "Classes of pebble games and complete problems," *SIAM J. Comput.*, vol. 8, no. 4, pp. 574–586, 1979.

[16] P. G. Kolaitis and J. Panttaja, "On the complexity of existential pebble games," in *Proc. CSL'03*, 2003, pp. 314–329.

[17] J. Nordström, "Pebble games, proof complexity, and time-space trade-offs," *Logical Methods in Computer Science*, 2012, to appear.

[18] J. Nordström and J. Håstad, "Towards an optimal separation of space and length in resolution," in *Proc. STOC'08*, 2008.

[19] P. Pudlak, "Proofs as games," *The American Mathematical Monthly*, vol. 107, no. 6, pp. pp. 541–550, 2000.

[20] A. Urquhart, "Width and size of regular resolution proofs," Talk given at the Banff Proof Complexity Workshop, 2011.